\documentclass[usenatbib]{mnras}

\usepackage{newtxtext,newtxmath}
\usepackage[T1]{fontenc}
\usepackage[utf8]{inputenc}
\usepackage{ae,aecompl}

\usepackage[usenames, dvipsnames]{color}
\usepackage{graphicx}	
\usepackage{amsmath}	
\usepackage{pifont}
\usepackage{multirow}
\usepackage[normalem]{ulem}
\usepackage{arydshln}

\newcommand{\km}{\ensuremath{\, \mathrm{km}}}
\newcommand{\pc}{\ensuremath{\, \mathrm{pc}}}

\newcommand{\Msun}{\ensuremath{\, \mathrm{M}_{\odot}}}
\newcommand{\Rsun}{\ensuremath{\, \mathrm{R}_{\odot}}}
\newcommand{\s}{\ensuremath{\, \mathrm{s}}}
\newcommand{\yr}{\ensuremath{\, \mathrm{yr}}}

\newcommand{\G}{\ensuremath{\, \mathrm{G}}}

\newcommand{\gal}{\ensuremath{\, \mathrm{gal}}}
\newcommand{\dex}{\ensuremath{\mathrm{\, dex}}}

\newcommand{\mo}{\ensuremath{^{-1}}}
\newcommand{\ie}{\emph{i.e.}\, }
\newcommand{\eg}{\emph{e.g.}\, }

\renewcommand{\d}{\mathrm{d}}

\newcommand{\xmark}{\ding{55}}

\newcommand{\Mbh}{\ensuremath{M_\bullet}}

\renewcommand{\vv}[1][args]{\mathbf}

\renewcommand{\log}{\mathrm{log}_{10}}

\title[TDE rate in nucleated galaxies]{Enhancement of the tidal disruption event rate in galaxies with a nuclear star cluster: from dwarfs to ellipticals}

\author[H. Pfister et al.]{
Hugo Pfister,$^{1,2}$\thanks{Sophie and Tycho Brahe Fellow; hugo.pfister@nbi.ku.dk}
Marta Volonteri,$^{3}$ Jane Lixin Dai$^{2,1}$ and Monica Colpi$^{4,5}$\\
$^{1}$DARK, Niels Bohr Institute, University of Copenhagen, Blegdamsvej 17, DK-2100 Copenhagen, Denmark\\
$^{2}$Department of Physics, The University of Hong Kong, Pokfulam Road, Hong Kong, China\\
$^{3}$Sorbonne Universit\'{e}, CNRS, UMR7095, Institut d'Astrophysique de Paris, 98bis boulevard Arago, F-75014, Paris, France\\
$^{4}$Dipartimento di Fisica G. Occhialini, Universit$\grave{a}$ degli Studi di Milano--Bicocca, Piazza della Scienza 3, I-20126 Milano, Italy\\
$^5$National Institute of Nuclear Physics INFN, Milano - Bicocca, Piazza della Scienza 3, 20126 Milano, Italy
}

\date{Accepted XXX. Received YYY; in original form ZZZ}

\pubyear{2019}

\begin{document}
\label{firstpage}
\pagerange{\pageref{firstpage}--\pageref{lastpage}}
\maketitle

\begin{abstract}
We compute the tidal disruption event (TDE) rate around local massive black holes (MBHs) with masses as low as $2.5\times10^4\Msun$, thus probing the dwarf regime for the first time. We select a sample of 37 galaxies for which we have the surface stellar density profile, a dynamical estimate of the mass of the MBH, and 6 of which, including our Milky Way, have a resolved nuclear star cluster (NSC). For the Milky Way, we find a total TDE rate of $\sim 10^{-4}\yr^{-1}$ when taking the NSC in account, and $\sim 10^{-7} \yr^{-1}$ otherwise. TDEs are mainly sourced from the NSC for light ($<3\times 10^{10}\Msun$) galaxies, with a rate of few $10^{-5}\yr^{-1}$, and an enhancement of up to 2 orders of magnitude compared to non-nucleated galaxies. We create a mock population of galaxies using different sets of scaling relations to explore trends with galaxy mass, taking into account the nucleated fraction of galaxies. Overall, we find a rate of few $10^{-5}\yr^{-1}$ which drops when galaxies are more massive than $10^{11}\Msun$ and contain MBHs swallowing stars whole and resulting in no observable TDE.
\end{abstract}


\section{Introduction}

When a star passes sufficiently close to a massive black hole (MBH), it can get accreted. For solar-type stars and MBHs with mass up to $\sim 10^8 \Msun$, the star is not swallowed whole, but it is tidally perturbed and destroyed, with a fraction of its mass falling back on to the MBH causing a bright flare, known as a tidal disruption event \citep[TDE;][]{Hills_75, Rees_88}.  As transient luminous events, TDEs are excellent candidates to discover low luminosity dormant intermediate mass black holes in dwarf galaxies \citep{Greene_19}. Moreover, as stars are not subject to feedback, which prevents MBH growth in dwarf galaxies \citep{Dubois_15, Trebitsch_18}, repeated TDEs, and subsequent accretion of stellar debris, could be a mechanism to ``grow'' these intermediate mass black holes \citep{Rees_88, Alexander_17}.

From an observational perspective, with a handful of observed TDEs, in the X-ray \citep[\eg][]{Auchettl_17} where most of the emission is produced, or in the optical/UV \citep[\eg][]{VanVelzen_11, Gezari_12, VanVelzen_20} for which surveys can cover a large area of the sky, estimating the TDE rate per galaxy starts becoming possible. {Different groups are converging towards a rate of $\sim10^{-4}\yr\mo\gal\mo$ \citep{VanVelzen_11, VanVelzen_18, Auchettl_18,Hung2018}, however, the exact case-by-case rate depends on the exact properties of the galaxy \citep{French_20b}: density profile, mass of the central MBH, stellar mass function, star formation rate etc.. For instance, galaxies which had a startburst about 1~Gyr ago and currently exhibit no sign of star formation (the E+A galaxies) appear to have a higher TDE rate \citep[they represent $\sim1\%$ of galaxies and host $\gtrsim 10\%$ of TDEs, \eg][]{French_16,LawSmith_17, Graur_18}. Similarly, \cite{Tadhunter_17} found a TDE in a rare ultraluminous infrared galaxy, suggesting that they could have a TDE rate as high as $10^{-1}\yr\mo\gal\mo$.}

{Unfortunately, the number of observed TDEs is still too low to slice the galaxy/BH plane in various properties, for instance \cite{VanVelzen_18} computed the TDE rate as a function of the galaxy/BH mass using 16 TDEs. However next generation facilities \citep[the LSST and eROSITA, \eg][]{VanVelzen_11, Jonker_20} will detect up to thousands of TDEs and make this ``slicing'' possible, allowing us to confront theoretical models for the dependency of the TDE rate with galaxy/BH properties.}

From a theoretical perspective, the most efficient way to bring stars close enough to the MBH to be disrupted is 2-body interactions \citep{Lightman_77, Merrit_book}. \cite{Wang_04} find that the TDE rate in an isothermal sphere surrounding a MBH lying on the $M_\bullet - \sigma\,$ relation \citep{Kormendy_13} is:
\begin{eqnarray}
\Gamma = 6.5\times 10^{-4}\yr^{-1} \left(  \frac{M_\bullet}{10^6 \Msun}\right)^{-0.25} \, ,
\end{eqnarray}
where $M_\bullet$ is the mass of the MBH. \cite{Stone_16a} find similar rates using a subsample of 144 observed galaxies from \cite{Lauer_07} for which the density profile is available, hence breaking the assumption of the isothermal sphere. Note that the assumption of the central MBH lying on the $M_\bullet - \sigma$ relation is still made. 

These two works suggest that lighter MBHs, \ie MBHs in dwarf galaxies, should exhibit a larger rate of TDEs. In addition, the $\Lambda$CDM paradigm predicts that dwarf galaxies are the most numerous in our Universe \citep{Bullock_17}. All this suggests that most TDEs should come from MBHs with masses $M_\bullet \lesssim 10^6\Msun$. {This is not what is found, with a clear drop of the \emph{observed} number of TDEs for MBHs with masses lower than $\sim 10^6 \Msun$ \citep{Wevers_19}. However, \cite{Wang_04} and \cite{Stone_16a} provide estimates of the rate at which stars are disrupted, which is \textit{a priori} different from the \textit{observable} TDE rate, as some TDEs may not be detected. Indeed, the observability of TDEs depends on additional physics \citep[\eg the overall debris mass supply rate determined by the mass and internal structure of the star, the circularisation efficiency determined by the stellar orbital parameters and black hole properties, the emission mechanism or dust obscuration; ][]{Kesden_12, Guillochon_13, Piran_15, Dai_15, Roth_16, Dai_18,  Mockler_19} and TDEs around lighter MBHs are fainter (considering the emission is capped by the Eddington luminosity).}

\begin{figure}
\includegraphics[width=\columnwidth]{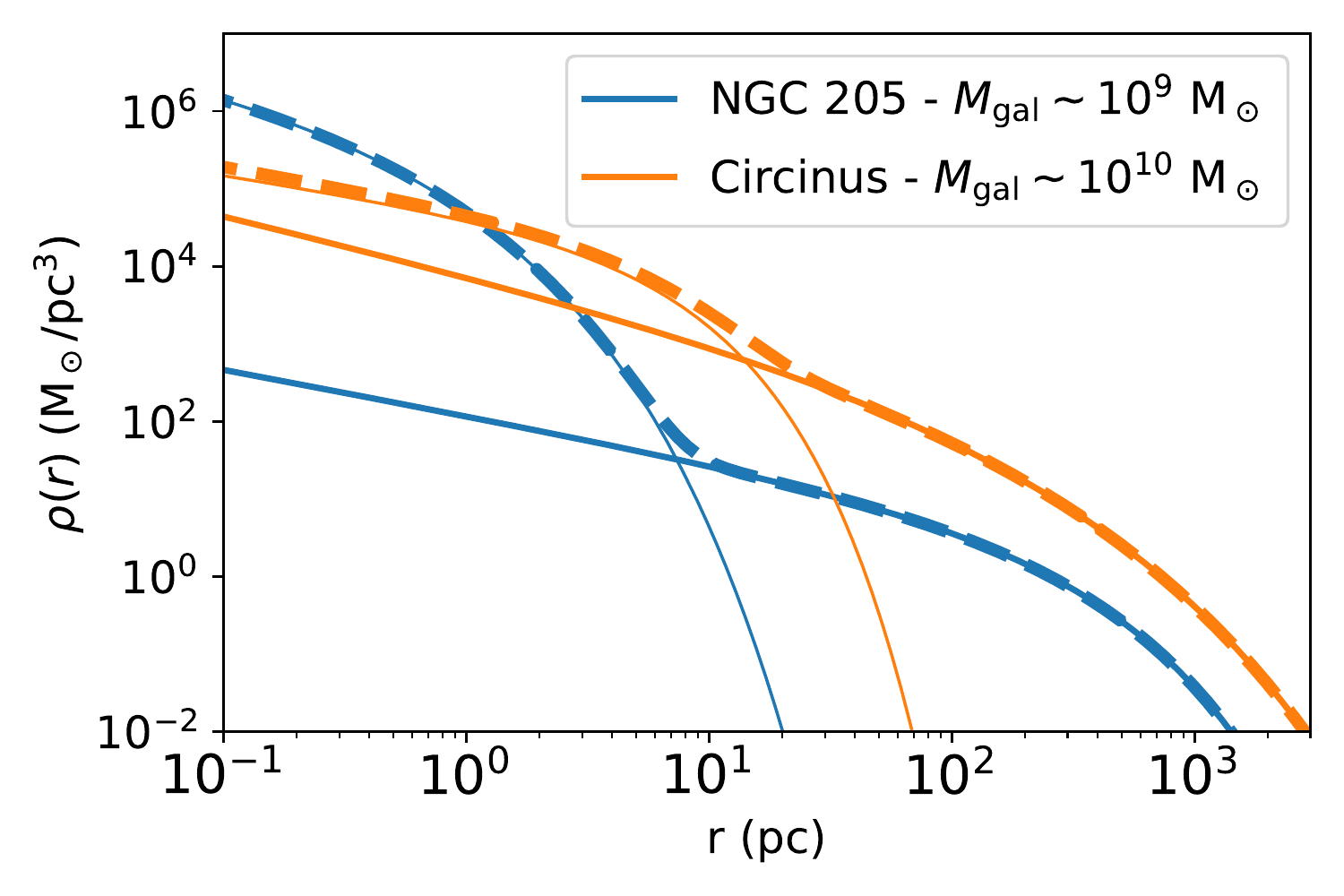}
\caption{Stellar density of the dwarf galaxy NGC~205 (blue) and Circinus (orange). The total density is shown with dashed lines, the density of the bulge with thick solid lines and the one of the NSC with thin solid lines. All quantities are shown as a function of the distance to the center of the galaxy. In the absence of NSCs, the central density would be lower by orders of magnitude.}
\label{fig:RhoOfr}
\end{figure}

{In addition, it could be that the assumptions made by \cite{Stone_16a} and \cite{Wang_04} break down at these low masses. As an example, their work assume that the central MBH lies on the $M_\bullet - \sigma$ relation, which is tightly constrained for MBHs with masses $\gtrsim 10^6 \Msun$, but exhibits a large scatter in the dwarf regime \citep{Greene_19}.} Furthermore, none of these previous works take into account that some galaxies may harbour a nuclear star cluster (NSC). The environments in the center of these nucleated galaxies differ significantly from those in non-nucleated galaxies. As an example, we show in Fig.~\ref{fig:RhoOfr} the total density profile (dashed lines) as well as the bulge/NSC (thick/thin solid lines) decomposition of the dwarf galaxy NGC~205 \citep[blue; ][]{Nguyen_18} and Circinus \citep[orange; ][]{Pechetti_19}.
In the absence of NSC, the central density in the dwarf NGC~205 would be up to 4 orders of magnitude lower. Not only the enhancement is lower in Circinus, but the fraction of nucleated galaxies is lower at higher mass \citep{SanchezJanssen_19}. The extreme density found in NSCs is known to speed up the formation of binary MBHs due to more efficient dynamical friction and stellar scattering \citep[\eg][]{Biava_19}, but also to boost the TDE rate \citep{MastrobuonoBattisti_14, Aharon_16,ArcaSedda_17}. All this suggests that contributions of NSCs in the dwarf regime should play a major role.

In this paper, we estimate the TDE rate for a sample of 37 galaxies (\S\ref{sec:ApplicationToRealData}) and for a mock catalog built using a set of scaling relations (\S\ref{sec:UnderstandingTheTrendWithMockCatalogs}). For these two samples, \textit{(i)} some MBHs have masses as low as few $10^4\Msun$ allowing us to study the TDE rate in the dwarf regime; \textit{(ii)} we relax the assumption that MBHs lie exactly on the $M_\bullet - \sigma$ relation; and \textit{(iii)} some galaxies have a NSC, allowing us to study the relative contribution of this component compared with the one of the bulge.

\section{TDE rate}
\label{sec:TDErate}

In this Section, we explain how we estimate the TDE rate (\S\ref{sec:EstimateOfTheTDERate}) given a density profile (\S\ref{sec:DensityProfile}).

\subsection{Estimate of the TDE rate}
\label{sec:EstimateOfTheTDERate}

We adopt an approach similar to \cite{Pfister_19b} to estimate the TDE rate. A spherical density profile $\rho(r)$ with a central MBH is provided as an input to \textsc{PhaseFlow} \citep{Vasiliev_17, Vasiliev_18}, which computes the following quantities:
\begin{itemize}
\item the stellar distribution function $f(E)$, which is further assumed to be ergodic, obtained through the Eddington inversion \citep{BT_87}. $E=v^2/2+\phi(r)$ is the energy per unit mass, $r$ and $v$ are respectively the distance to the center and relative speed, and $\phi$ is the galactic gravitational potential. Once $f$ is estimated, we verify if is positive everywhere;
\item the energy density function $N(E)=4\pi^2L^2_c(E)f(E)P(E)$ \citep{Merrit_book}. $L_c(E)$ and $P(E)$ represent respectively the circular angular momentum and radial period of stars with energy $E$;
\item  the loss-cone filling factor $q(E)$  \citep[Eq.~(13a) from ][]{Vasiliev_17};
\item the loss-cone boundary $\mathcal{R}_\mathrm{LC}$  \citep[Eq.~(13b) from ][]{Vasiliev_17};
\item the orbit-averaged diffusion coefficient $\mu$   \citep[Eq.~(13c) from ][]{Vasiliev_17};
\end{itemize}

With this information, we can infer the flux of stars entering the loss-cone region ($\mathcal{F}$) per unit energy (Eq.~(16) from \cite{Stone_16a}, Eq.~(14) from \cite{Vasiliev_17}, Eq.~(8) from \cite{Pfister_19b}):
\begin{eqnarray}
\frac{\partial \mathcal{F}}{\partial E}= \frac{q(E)\, \mathcal{R}_\mathrm{LC}}{(q(E)^2+q(E)^4)^{1/4}+\ln(1/\mathcal{R}_\mathrm{LC})} \frac{N(E)}{P(E)}.
\end{eqnarray}
{This equation can be integrated to obtain the total flux of stars entering the loss-cone region, in units of number of stars per year per galaxy. We make the assumption that this flux equals the TDE rate, \ie all stars penetrating the loss-cone region result in a TDE (but for MBHs with $M_\bullet \geq 10^8 \Msun$, see below), so that $\Gamma=\mathcal{F}$. While this not a terrible approximation, additional physics is needed to determine if the TDE is \textit{observable}: stars being disrupted by light ($\lesssim 10^6\Msun$) MBHs may result in a faint event (if the emission is capped by the Eddington luminosity) that are less likely to be observed with current facilities, and stars being disrupted by massive MBHs ($\gtrsim 10^8\Msun$) may be swallowed whole resulting in no flare \citep[\eg ][]{Rees_88}. In addition, we assume that the stellar population is monochromatic and solar like, so that stars all have the same mass ($m_\star=\Msun$) and radius ($r_\star = \Rsun$). While this assumption compared with a  non-monochromatic population only changes the flux of stars entering the loss-cone region by a factor of $\sim 2$ \citep{Stone_16a}, this probably changes the \textit{observable} TDE rate, as most TDEs are sourced by sub-solar mass stars \citep{Kroupa_01, Mockler_19} resulting in fainter events. Finally, some stars entering in the loss-cone region may only be partially disrupted \citep{Mainetti2017}, also resulting in fainter events: \cite{Guillochon_13} found that stars with polytropic index $\eta=5/3$ entering the loss-cone boundary are fully disrupted, but only $\sim5\%$ of the mass is lost for $\eta=4/3$ stars (or equivalently, the loss-cone region for full disruption of $\eta=4/3$ stars is smaller than for $\eta=5/3$ stars).}

{In order to take into account that the most massive MBHs would swallow stars whole resulting in no TDE, we consider that MBHs with $M_\bullet \geq M_{H,\odot}=10^8\Msun$ have a null TDE rate. In reality this threshold depends on the spin of the MBH and on the mass of the star \citep{Ivanov2006, Kesden2012,Stone2020} yielding limiting values for the MBH mass for direct capture between $10^7\Msun<M_\bullet<10^9\Msun,$ with the lowest value for a star of $0.1\Msun$ around a non-spinning MBH and the highest for a massive star around a highly spinning MBH. For a real non-monochromatic population of stars, this results in a smooth transition starting at $M_\bullet\sim10^7\Msun$ and ending around $M_\bullet\sim10^9\Msun$ rather than a sharp cut. We note that only one (possible) \textit{observed} TDE has been associated to a MBH with mass larger than this \citep[ASASSN-15lh could be powered by a $>10^{8.3}\Msun$ MBH; ][]{Dong_16, Kruhler_18, Mummery_20}.}

{These additional physical processes affecting the observability of TDEs are beyond the scope of this study. Throughout this paper, we refer to the TDE rate as the rate at which stars get close enough to the MBH to be disrupted without being swallowed whole. This TDE rate is therefore an upper limit for the local \textit{observable} TDE rate which we may detect.}

\subsection{Density profiles}
\label{sec:DensityProfile}

Surface density profiles are often \citep[\eg][]{Lauer_07,SanchezJanssen_19} fitted with a Sérsic profile \citep{Sersic_68} which depends on 3 parameters: the mass of the structure $M_\star$, the Sérsic index $n$, and the effective radius $R_\mathrm{eff}$\footnote{We parametrize the Sérsic profile so that the effective radius is equal to the half-light radius.}. It has been shown by \cite{Prugniel_97} and \cite{Marquez_00} that the underlying three dimensional density profile, which we will refer to as the Prugniel profile throughout this paper, is well approximated by:
\begin{eqnarray}
\rho(r) &=& \rho_0 \left( \frac{r}{R_\mathrm{eff}} \right)^{-p} e^{-b \left(r/R_\mathrm{eff}\right)^{1/n}} \label{eq:Spheroid}\\
p &=& 1 - \frac{0.6097}{n}+ \frac{0.05563}{n^2} \label{eq:p}\\
b &=& 2 n-\frac{1}{3}+\frac{0.009876}{n} \label{eq:b}\\ 
\rho_0 &=& \frac{M_\star}{4\pi R^3_\mathrm{eff}} \times \frac{b^{n(3-p)}}{n\times \gamma_\mathrm{E}(n(3-p))} \label{eq:rho0}  \, ,
\end{eqnarray}
where Eq.~\eqref{eq:rho0} comes from mass conservation and $\gamma_\mathrm{E}$ is the Euler Gamma function\footnote{$\gamma_\mathrm{E}(x)=\int_0^\infty t^{x-1}e^{-t}\d t$}. We note that our Sérsic parametrisation only allows for fairly flat inner 3D logarithmic slope ($p<1$). While this may result in under-estimates of the TDE rate \citep[see Fig.~5 of ][]{Stone_16a}, this is motivated observationally as the Sérsic profile is widely used and accurately fits the observed luminosity 2D profiles of galaxies as well as NSCs \citep[\eg ][but an example can be found in Fig.~\ref{fig:RhoOfr}]{SanchezJanssen_19}.\\

Our strategy is therefore the following: for a given structure, \ie a galaxy, a bulge, or a NSC,  with surface density  fitted with a Sérsic profile, we reconstruct the associated three dimensional Prugniel density profile (Eq.~(\ref{eq:Spheroid}-\ref{eq:rho0})) and add a central MBH with mass $M_\bullet$. From this, the TDE rate can be estimated as explained in \S\ref{sec:EstimateOfTheTDERate}.

\section{Application to real galaxies}
\label{sec:ApplicationToRealData}

\begin{figure*}
\includegraphics[width=\columnwidth]{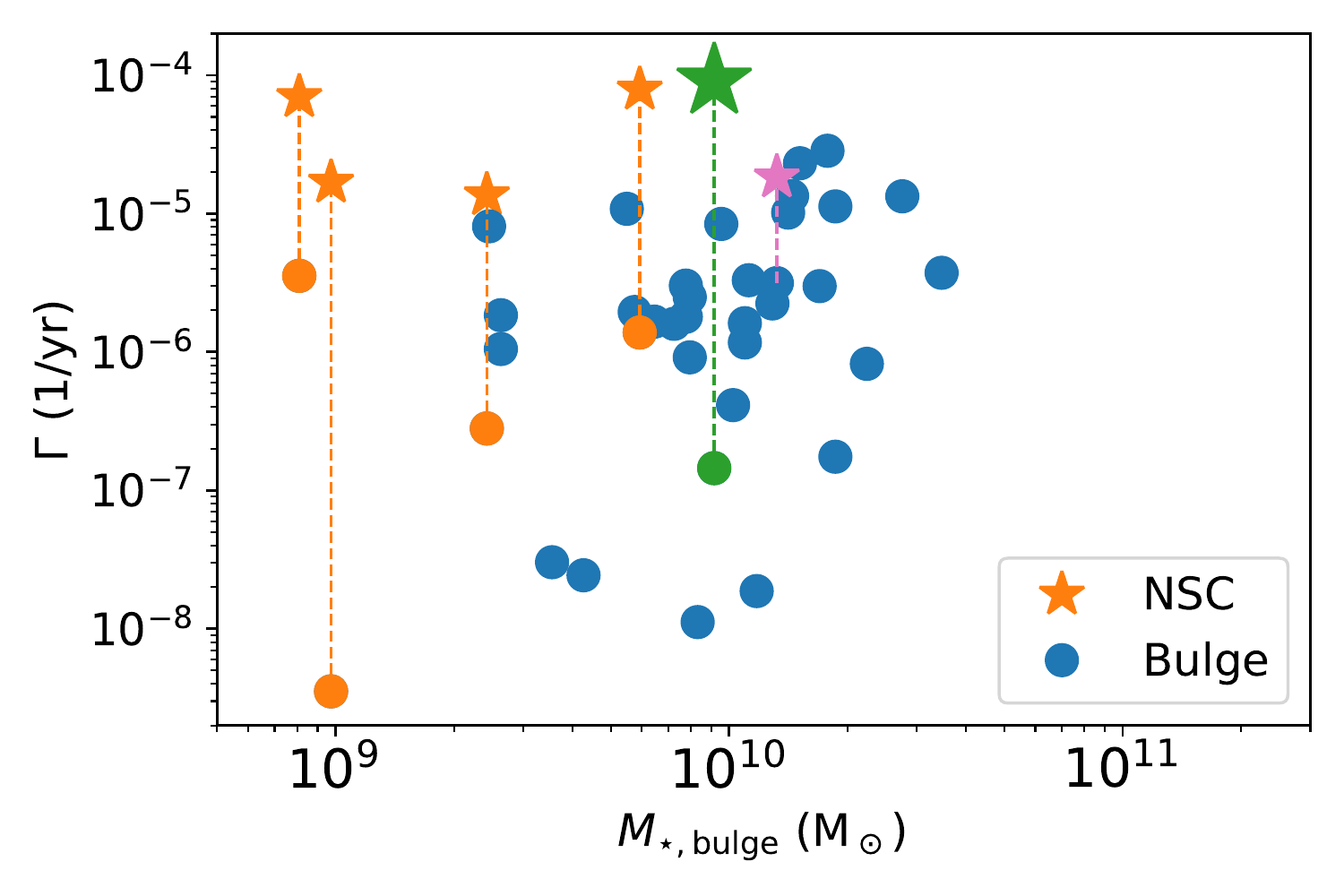} 
\hfill
\includegraphics[width=\columnwidth]{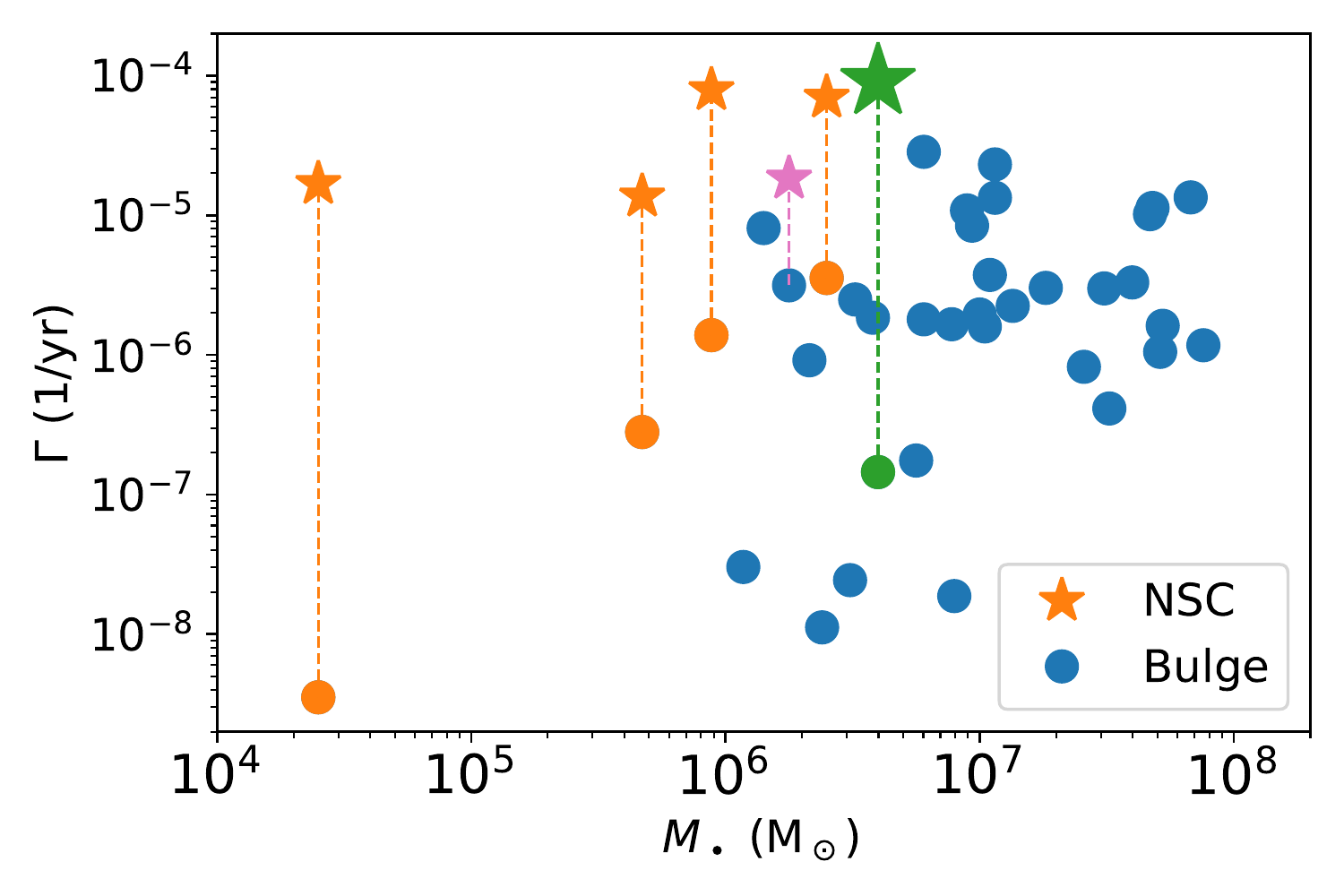} 
\caption{\textbf{Left:} TDE rate as a function of the mass of the bulge. \textbf{Right:} TDE rate as a function of the mass of the MBH. In both cases we show the TDE rate sourced from the bulge (circles) which can be interpreted as the TDE rate one would infer using {observations of galaxies with unresolved NSC}, and from the NSC (stars). The different colors indicate where we obtained the data: {\protect \cite{Nguyen_18}} in orange; {\protect \cite{Davis_19}} in dark blue; {\protect \cite{Pechetti_19}} in pink and the Milky Way using the NSC properties of {\protect \cite{Schodel_18}} in green. Overall, there is a large scatter and no trend clearly appears, but galaxies with a NSC see their TDE rate enhanced by $\sim$2 orders of magnitude when it is taken into account.}
\label{fig:GammaObs}
\end{figure*}

In this Section, we apply the technique described in \S\ref{sec:EstimateOfTheTDERate} to real galaxies to obtain the TDE rate. We describe the data we use in \S\ref{sec:Data}, give our results in \S\ref{sec:Rates} and a possible interpretation in \S\ref{sec:FastEstimate}.

\subsection{Data}
\label{sec:Data}

\subsubsection{``Unresolved'' galaxies}

\cite{Davis_19} published a list of 40 galaxies (including the Milky Way) hosting a MBH, for which they provide $M_{\star,\mathrm{bulge}},\, n_\mathrm{bulge},\, R_{\mathrm{eff},\mathrm{bulge}}$ of the bulge and a dynamical (\ie not assuming the $M_\bullet-\sigma$ relation) estimate of $M_\bullet$.

\subsubsection{``Resolved'' galaxies}
\label{sec:ResolvedGalaxies}

Similarly to \cite{Biava_19}, we use the data of \cite{Nguyen_18} who published a study of four galaxies hosting a MBH. For all of their galaxies, they provided the Sérsic quantities $n_\mathrm{bulge}$, $M_{\star,\mathrm{bulge}}$, $R_{\mathrm{eff},\mathrm{bulge}}$ of the bulge, a dynamical estimate of the mass of the MBH $M_\bullet$, as well as the Sérsic quantities $n_\mathrm{NSC}$, $M_{\star, \mathrm{NSC}}$, $R_{\mathrm{eff, NSC}}$ of the central NSC. In addition, the recent paper of \cite{Pechetti_19} provides additional Sérsic fits for 29 NSCs, 2 of which belongs to galaxies (Circinus and NGC~5055) included in the sample of \cite{Davis_19}. 

\subsubsection{The Milky Way}

Particular care is taken for the Milky Way. \cite{Davis_19} provides the Sérsic parameters for the bulge. Regarding the NSC, we fit the observed luminosity profile of the inner pc of our Galaxy \citep[Fig.~9 of][]{Schodel_18} with a Sérsic profile. We obtain $(R_\mathrm{eff, NSC}/\pc, n_\mathrm{NSC})\sim(6,2)$. We choose $M_{\star,\mathrm{NSC}}=4.4\times 10^7 \Msun$ so that the density at 1 pc (0.1 pc) is $1.6\times10^5\Msun  \pc^{-3}$ ($2.2\times10^6 \Msun \pc^{-3}$) and the mass within 1 pc is $1.3\times10^6\Msun$, in agreement with the value given in Table~3 of \cite{Schodel_18}.\newline

We remove galaxies for which the mass of the MBH is larger than $10^8\Msun$, in order to take into account that no TDE would be seen in this situation. This reduces the ``observed'' sample to 37 galaxies, 6 of which, including our Milky Way, have a resolved NSC (see Table~\ref{table:catalog}). For all these galaxies we use the method described in \S\ref{sec:EstimateOfTheTDERate} to obtain the TDE rate. For the 6 galaxies with a resolved NSC, we compute separately the TDE rate for stars in the bulge and stars in the NSC. This allows us to study the relative contribution of each component, and the total TDE rate is simply obtained summing the TDE rate from all components. Our results can be found in Table~\ref{table:catalog}.

\begin{center}
\begin{table*}
\begin{tabular}{ccccccccc}
\hline
Name & Component & $\log \left( \frac{M_\star}{\Msun} \right)$ & $n$ &   $\log \left( \frac{R_\mathrm{eff}}{\pc} \right)$ & $\log \left( \frac{M_\bullet}{\Msun} \right)$ & $\log \left( \frac{\rho_\mathrm{inf}}{\Msun \pc^{-3}} \right)$& $\log \left(\frac{\Gamma}{\yr\mo} \right)$ & Source \\
\hline
\hline
Milky Way &       bulge &      9.96 &  1.30 &     3.02 &    6.60 &       1.29 &   -6.84 &     \cite{Davis_19} \\
 &       NSC &      7.64 &  2.00 &     0.78 &    6.60 &       5.98 &   -4.04 &   \cite{Schodel_18} \\
Circinus &       bulge &     10.12 &  2.21 &     2.83 &    6.25 &       3.43 &   -5.50 &     \cite{Davis_19} \\
 &       NSC &      7.57 &  1.09 &     0.90 &    6.25 &       4.60 &   -4.74 &  \cite{Pechetti_19} \\
M32 &       bulge &      8.90 &  1.60 &     2.03 &    6.40 &       3.40 &   -5.45 &    \cite{Nguyen_18} \\
 &       NSC &      7.16 &  2.70 &     0.64 &    6.40 &       6.46 &   -4.16 &    \cite{Nguyen_18} \\
NGC 205 &       bulge &      8.99 &  1.40 &     2.71 &    4.40 &       1.67 &   -8.45 &    \cite{Nguyen_18} \\
 &       NSC &      6.26 &  1.60 &     0.11 &    4.40 &       6.36 &   -4.78 &    \cite{Nguyen_18} \\
NGC 5102 &       bulge &      9.77 &  3.00 &     3.08 &    5.94 &       3.33 &   -5.86 &    \cite{Nguyen_18} \\
 &       NSC &      6.85 &  0.80 &     0.20 &    5.94 &       5.67 &   -4.21 &    \cite{Nguyen_18} \\
 &       NSC &      7.76 &  3.10 &     1.51 &    5.94 &       5.34 &   -4.76 &    \cite{Nguyen_18} \\
NGC 5206 &       bulge &      9.38 &  2.57 &     2.99 &    5.67 &       2.62 &   -6.55 &    \cite{Nguyen_18} \\
 &       NSC &      6.23 &  0.80 &     0.53 &    5.67 &       4.16 &   -5.66 &    \cite{Nguyen_18} \\
 &       NSC &      7.11 &  2.30 &     1.02 &    5.67 &       5.15 &   -4.94 &    \cite{Nguyen_18} \\
ESO 558-G009 &       bulge &      9.89 &  1.28 &     2.52 &    7.26 &       2.53 &   -5.52 &     \cite{Davis_19} \\
IC 2560 &       bulge &      9.63 &  2.27 &     3.62 &    6.49 &       0.40 &   -7.61 &     \cite{Davis_19} \\ 
J0437+2456 &       bulge &      9.90 &  1.73 &     2.62 &    6.51 &       3.04 &   -5.60 &     \cite{Davis_19} \\
Mrk 1029 &       bulge &      9.90 &  1.15 &     2.48 &    6.33 &       2.65 &   -6.04 &     \cite{Davis_19} \\
NGC 0253 &       bulge &      9.76 &  2.53 &     2.97 &    7.00 &       2.68 &   -5.71 &     \cite{Davis_19} \\
NGC 1068 &       bulge &     10.27 &  0.71 &     2.71 &    6.75 &       1.51 &   -6.76 &     \cite{Davis_19} \\
NGC 1320 &       bulge &     10.25 &  3.08 &     2.79 &    6.78 &       4.65 &   -4.55 &     \cite{Davis_19} \\
NGC 2273 &       bulge &      9.98 &  2.24 &     2.66 &    6.97 &       3.56 &   -5.07 &     \cite{Davis_19} \\
NGC 2960 &       bulge &     10.44 &  2.59 &     2.91 &    7.06 &       3.84 &   -4.87 &     \cite{Davis_19} \\
NGC 3031 &       bulge &     10.16 &  2.81 &     2.79 &    7.83 &       3.79 &   -4.87 &     \cite{Davis_19} \\
NGC 3079 &       bulge &      9.92 &  0.52 &     2.67 &    6.38 &       0.85 &   -7.95 &     \cite{Davis_19} \\
NGC 3227 &       bulge &     10.04 &  2.60 &     3.26 &    7.88 &       1.94 &   -5.93 &     \cite{Davis_19} \\
NGC 3368 &       bulge &      9.81 &  1.19 &     2.49 &    6.89 &       2.45 &   -5.78 &     \cite{Davis_19} \\
NGC 3393 &       bulge &     10.23 &  1.14 &     2.63 &    7.49 &       2.34 &   -5.52 &     \cite{Davis_19} \\
NGC 3627 &       bulge &      9.74 &  3.17 &     2.76 &    6.95 &       4.05 &   -4.97 &     \cite{Davis_19} \\
NGC 4151 &       bulge &     10.27 &  2.24 &     2.76 &    7.68 &       3.43 &   -4.95 &     \cite{Davis_19} \\
NGC 4258 &       bulge &     10.05 &  3.21 &     3.19 &    7.60 &       2.97 &   -5.48 &     \cite{Davis_19} \\
NGC 4303 &       bulge &      9.42 &  1.02 &     2.15 &    6.58 &       2.82 &   -5.73 &     \cite{Davis_19} \\
NGC 4388 &       bulge &     10.07 &  0.89 &     3.27 &    6.90 &      -0.07 &   -7.73 &     \cite{Davis_19} \\
NGC 4501 &       bulge &     10.11 &  2.33 &     3.06 &    7.13 &       2.59 &   -5.65 &     \cite{Davis_19} \\
NGC 4736 &       bulge &      9.89 &  0.93 &     2.32 &    6.78 &       2.66 &   -5.75 &     \cite{Davis_19} \\
NGC 4826 &       bulge &      9.55 &  0.73 &     2.57 &    6.07 &       1.25 &   -7.52 &     \cite{Davis_19} \\
NGC 4945 &       bulge &      9.39 &  3.40 &     2.67 &    6.15 &       4.40 &   -5.09 &     \cite{Davis_19} \\
NGC 5495 &       bulge &     10.54 &  2.60 &     3.25 &    7.04 &       2.98 &   -5.43 &     \cite{Davis_19} \\
NGC 5765 &       bulge &     10.04 &  1.46 &     2.86 &    7.72 &       1.84 &   -5.79 &     \cite{Davis_19} \\
NGC 6264 &       bulge &     10.01 &  1.04 &     2.92 &    7.51 &       1.06 &   -6.38 &     \cite{Davis_19} \\
NGC 6323 &       bulge &      9.86 &  2.09 &     2.92 &    7.02 &       2.42 &   -5.80 &     \cite{Davis_19} \\
NGC 7582 &       bulge &     10.15 &  2.20 &     2.71 &    7.67 &       3.37 &   -4.99 &     \cite{Davis_19} \\
UGC 3789 &       bulge &     10.18 &  2.37 &     2.58 &    7.06 &       4.20 &   -4.64 &     \cite{Davis_19} \\
UGC 6093 &       bulge &     10.35 &  1.55 &     3.13 &    7.41 &       1.59 &   -6.09 &     \cite{Davis_19} \\
\hdashline
NGC 5055 &       bulge &     10.49 &  2.02 &     3.37 &    8.94 &       1.21 &    \xmark &     \cite{Davis_19} \\
 &       NSC &      7.71 &  2.75 &     1.16 &    8.94 &        \xmark &         \xmark &  \cite{Pechetti_19} \\
Cygnus A &       bulge &     12.36 &  1.45 &     4.34 &    9.44 &      -0.17 &    \xmark &     \cite{Davis_19} \\
NGC 4594 &       bulge &     10.81 &  6.14 &     3.32 &    8.81 &       6.07 &    \xmark &     \cite{Davis_19} \\
NGC 4699 &       bulge &     11.12 &  5.35 &     3.45 &    8.34 &       5.68 &    \xmark &     \cite{Davis_19} \\
NGC 2974 &       bulge &     10.23 &  1.56 &     2.98 &    8.23 &       1.71 &    \xmark &     \cite{Davis_19} \\
NGC 1398 &       bulge &     10.57 &  3.44 &     3.32 &    8.03 &       3.37 &    \xmark &     \cite{Davis_19} \\
NGC 1097 &       bulge &     10.83 &  1.95 &     3.28 &    8.38 &       2.02 &    \xmark &     \cite{Davis_19} \\
NGC 0224 &       bulge &     10.11 &  2.20 &     3.18 &    8.15 &       1.74 &    \xmark &     \cite{Davis_19} \\
         \hline
\end{tabular}
\caption{Full sample of observed galaxies with their inferred TDE rate. Galaxies below the horizontal dashed line have a MBH with a mass larger than $10^8\Msun$, therefore have a null TDE rate. For NGC~5206 and NGC~5102, which have two NSCs, we sum their contribution to estimate the TDE rate from NSCs, noting however that the TDE rate of the NSC with the higher $\rho_\mathrm{inf}$ dominates. We cannot compute $\rho_\mathrm{inf}$ for the NSC of NGC~5055 as it is more massive than the central MBH.}
\label{table:catalog}
\end{table*}
\end{center}

This sample is smaller than the one used by \cite{Stone_16a} to perform a similar analysis, but \textit{(i)} all the galaxies we consider have a dynamical estimate of the MBH mass, and we do not need to assume the MBH lies on the $M_\bullet - \sigma$ relation \citep[or the $M_\bullet - M_{\star,\mathrm{bulge}}$; ][]{Kormendy_13}; \textit{(ii)} we removed MBHs for which no TDE would be seen; \textit{(iii)} some of our galaxies have a resolved NSC; and \textit{(iv)} we extend the analysis to the dwarf galaxy regime, of crucial importance for both TDEs and gravitational wave studies with LISA \citep{LISA2017}.

\subsection{Rates}
\label{sec:Rates}

We show in Fig.~\ref{fig:GammaObs} the TDE rate generated by stars in the bulge (circles) and, for galaxies which have a NSC, the TDE rate generated by stars in the NSC (stars) as a function of the mass of the bulge (left panel) and of the MBH (right panel). The different colors indicate where we obtained the data (see caption).

We begin with the TDE rates originating from stars in bulges (circles), which we interpret as the TDE rate one would infer using a density profile obtained with observations of galaxies with unresolved NSCs. The size of a NSC is typically of 1-50~pc, corresponding to $\sim0.1''$ at 10~Mpc, thus NCSs would be unresolved in most galaxies \citep{Lauer_98, Stone_16b, SanchezJanssen_19,Pechetti_19}. Given the small number of MBHs at the low mass end ($M_\mathrm{\star,bulge}<5\times10^9\Msun$ and $M_\bullet < 10^6\Msun$), inferring ``trends'' would be dangerous; for the whole sample we find a mean TDE rate of $5\times10^{-6} \yr\mo$ and in general no significant dependence on bulge or MBH mass.

While this mean value is lower than current estimates, it does not take into account that some galaxies host a NSC in their center, which can enhance the TDE rate by orders of magnitude. Consider for instance the Milky Way, we find a TDE rate of $9.1\times 10^{-5} \yr^{-1}$ including the NSC (green star) and $1.4\times10^{-7} \yr^{-1}$ without it (green circle), resulting in an enhancement of $\sim 600$. This example shows how crucial it is to take into account NSCs when they exist. For the 6 galaxies for which the NSC is resolved and the density profiles is known, we find a total enhancement of the TDE rate when including NSCs varying between 6~(Circinus) to 4800~(NGC 205), with an average at 900. The mean TDE rate for these 6 NSCs is $5\times10^{-5}\yr^{-1}$.

This analysis illustrates how important it is to properly resolve NSCs to have a correct estimate of the TDE rate, as their presence/absence drastically changes the central density, changing the estimates of the TDE rate by orders of magnitude. However, all our lower mass MBHs are surrounded by a NSC, and conversely, none of our massive ones are. This is expected: the nucleation fraction has a peak of about 80--100\% for $10^9\Msun$ galaxies and decreases at lower and higher masses \citep{SanchezJanssen_19}; to assess more thouroughly the role of NSCs in sourcing TDEs, ideally we would need MBH mass measurements in a large sample of galaxies with and without resolved NSCs. Given that such observational sample is not available, in \S\ref{sec:UnderstandingTheTrendWithMockCatalogs}, we build a mock catalog of galaxies to perform this analysis.

\subsection{{A fast estimate of the TDE rate}}
\label{sec:FastEstimate}

\begin{figure}
\includegraphics[width=\columnwidth]{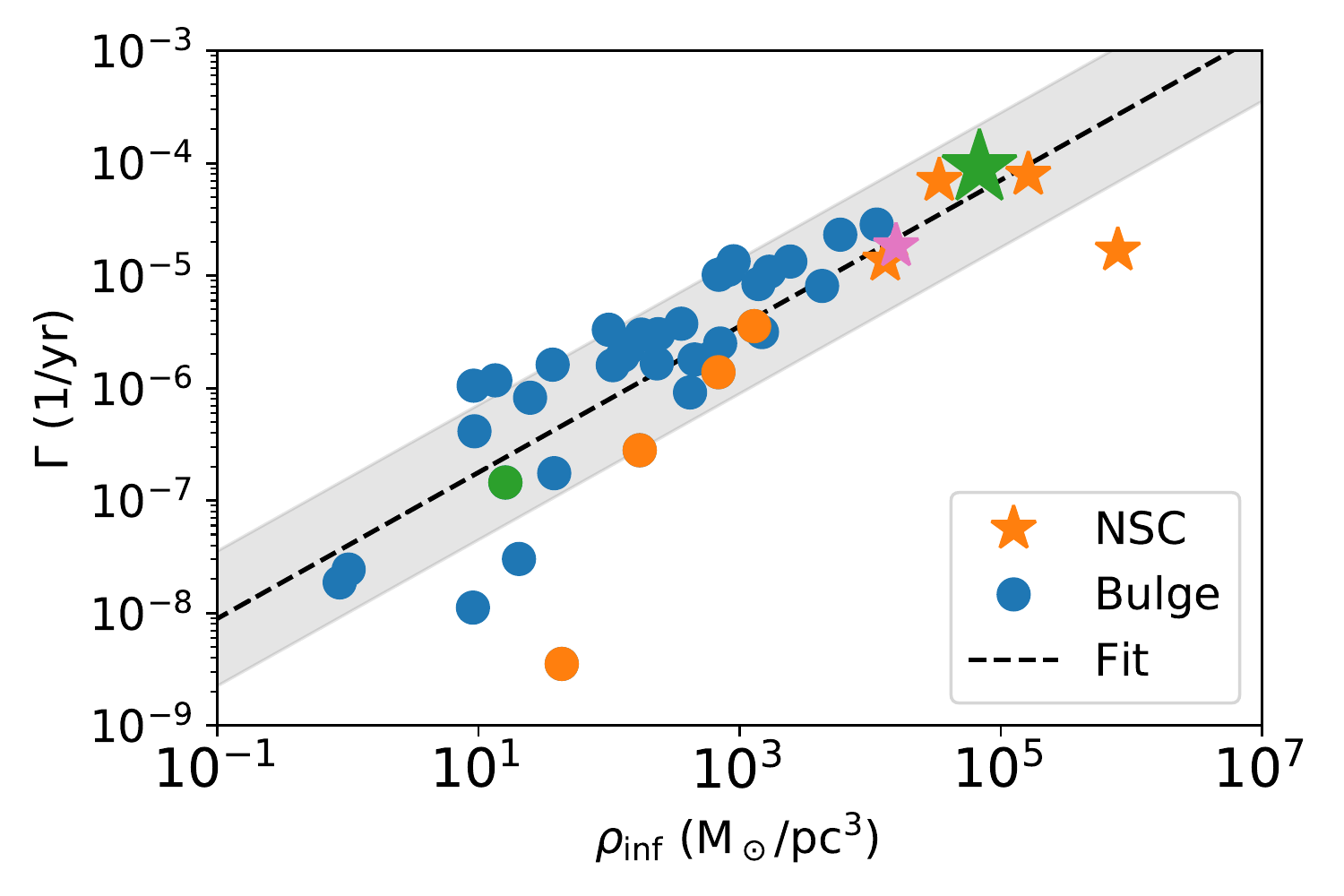}
\caption{TDE rate as a function of the density at the MBH influence radius for NSCs (stars) and bulges (circles). The different colors indicate where we obtained the data: {\protect \cite{Nguyen_18}} in orange; {\protect \cite{Davis_19}} in dark blue; {\protect \cite{Pechetti_19}} in pink and the Milky Way using the NSC of {\protect \cite{Schodel_18}} in green. We also indicate with the dashed black line the fit from Eq.~\eqref{eq:Gammaofrhoinf} as well as the 1-sigma scatter (0.6~dex) about the fit (shaded area).}
\label{fig:GammaVSrhoinf}
\end{figure}

Ideally, one would want to compute the TDE rate given the observed Sérsic properties of the structures ($M_\star$, $R_\mathrm{eff}$ and $n$) and the mass of the central MBH ($M_\bullet$) using a simple scaling. The TDE rate in a galaxy is in general a complex function of the four parameters describing the system in our model, however, a natural quantity tracing the TDE rate is the density at the gravitational influence radius ($\rho_\mathrm{inf}=\rho(r_\mathrm{inf})$), where the influence radius ($r_\mathrm{inf}$) is the radius at which the enclosed stellar mass is equal to that of the MBH. $\rho_\mathrm{inf}$ can be easily obtained from the properties of the MBH and the surrounding stellar structure, for instance for a Sérsic profile:
\begin{eqnarray}
r_\mathrm{inf}=\left\lbrace\frac{ \gamma^\mathrm{inv}_\mathrm{E,inc}\left[n(3-p); \frac{M_\bullet}{M_\star} \times \gamma_\mathrm{E}(n(3-p))\right]}{b} \right\rbrace ^n R_\mathrm{reff} \label{eq:r_inf} \, ,
\end{eqnarray}
where $\gamma^\mathrm{inv}_\mathrm{E,inc}(x;z)$ is the inverse of the incomplete Euler Gamma function\footnote{If the incomplete Euler Gamma function is $\gamma_\mathrm{E,inc}(x;y)=\int_0^y t^{x-1}e^{-t}\d t=z$, then $\gamma^\mathrm{inv}_\mathrm{E,inc}(x;z)=y$.}.

For the sample described in Section~\ref{sec:Data}, we show in Fig.~\ref{fig:GammaVSrhoinf} the TDE rate generated by stars in the bulge (circles) and, for galaxies which have a NSC, the TDE rate generated by stars in the NSC (stars) as a function of $\rho_\mathrm{inf}$. The different colors indicate where we obtained the data (see caption). 

In this situation, a clear trend arises, with larger $\rho_\mathrm{inf}$ resulting in larger TDE rates. This results is not surprising: the TDE flux peaks around the critical radius, corresponding to the radius at which the TDE flux of the full and empty loss-cone are equal, and the critical radius happens to be similar to the influence radius \citep{Syer_99,Wang_04,Merrit_book}. We fit this relation using a least-square regression in the $\ln\Gamma-\ln\rho_\mathrm{inf}$ plane and find:
\begin{eqnarray}
\frac{\Gamma}{\yr\mo}= 10^{-7.4\pm0.2} \left( \frac{\rho_\mathrm{inf}}{\Msun\pc^{-3}} \right)^{0.65\pm0.07} \label{eq:Gammaofrhoinf} \, ,
\end{eqnarray}
with a variance of 0.6~dex. 

The scaling of $\Gamma$ with $\rho_\mathrm{inf}$ explains why we found a higher TDE rate when an NSC is present: for the same MBH, the presence of an NSC implies a much higher stellar density near the MBH. Ignoring the presence of the NSC in these galaxies would lead to a large underestimate of the TDE rate. We will discuss further the relative importance of NSCs and bulges for different ranges of $M_\bullet$ and $M_\mathrm{\star, bulge}$ in Section~\ref{sec:UnderstandingTheTrendWithMockCatalogs}. 

The influence radius (hence the density at the influence radius) can be defined for any kind of density profile, for instance, in the case of a singular isothermal sphere with velocity dispersion $\sigma$, we have $\rho_\mathrm{inf}=2\sigma^6/ \pi \G^3 M^2_\bullet$, resulting in:
\begin{eqnarray}
\frac{\Gamma}{\yr\mo}= 3\times 10^{-4} \left( \frac{\sigma}{70\km\s\mo} \right)^{3.9} \left( \frac{M_\bullet}{10^6\Msun} \right)^{-1.3} \, .
\end{eqnarray}
This expression is in good agreement with Eq.~(29) of \cite{Wang_04} who find $\Gamma~=~7~\times~10^{-4}\yr\mo~(\sigma/70\km\s\mo)^{7/2}~(M_\bullet/10^6\Msun)^{-1}$. 

We recall that we obtained this expression for MBHs in Sérsic structures which have fairly flat inner 3D logarithmic slope ($p<1$), extrapolated it to a singular isothermal sphere with inner logarithmic slope $p=2$ and found a good agreement with previous analytical results. This suggests that this expression, which provides a rapid way to estimate the TDE rate without going through \textsc{PhaseFlow}, can be applied to a variety of density profiles with a central MBH.

\section{Understanding trends with mock catalogs}
\label{sec:UnderstandingTheTrendWithMockCatalogs}

In \S\ref{sec:ApplicationToRealData} we found a large scatter in the TDE rate as a function of MBH and bulge mass and no trend clearly arose from the small sample analyzed. This could be because a MBH with given $M_\bullet$ can be surrounded by a variety of structures, resulting in an intrinsically large variety of rates, or because the sample used is too small to highlight trends.

In this Section we perform a similar analysis but on a mock catalog based on empirical scaling relations for galaxies and MBHs. We describe how we build the catalog in \S\ref{sec:MockCatalog} and give our results in \S\ref{sec:Results}. This approach is useful in order to make statistical predictions for population of galaxies, and compare with large upcoming observational samples of TDEs.

\subsection{Mock catalog}
\label{sec:MockCatalog}

To produce a large realistic sample of galaxies from empirical relations, we proceed as following 100,000 times:
\begin{enumerate}
\item We draw a galaxy with total stellar mass $M_\mathrm{gal, star}$ from a log-uniform distribution between $10^9$ and $10^{12} \Msun$;
\item We compute the mass of the bulge $M_{\star,\mathrm{bulge}}$ fitting the median value of the ratio bulge to total mass \citep[Fig.~3 of][]{Kochfar_11}:
\begin{eqnarray}
M_{\star,\mathrm{bulge}} = M_\mathrm{gal,star} \times \mathrm{min} \left( 1 ; 10^{-10.1+0.9 \log{\left( {M_\mathrm{gal,star} / \Msun} \right) } }\right) \, ;
\end{eqnarray}

\item We compute the effective radius of the bulge $R_{\mathrm{eff,bulge}}$ using Eq.~(4) of \cite{Dabringhausen_08} (using all objects, read \S 3.1 of their paper):
\begin{eqnarray}
R_{\mathrm{eff}} = (2.95\pm 0.24)\pc \left(\frac{M_{\star}}{10^6\Msun}\right)^{0.596\pm 0.007} \, ; \label{eq:Dabringausen}
\end{eqnarray}
\item We compute the mass of the MBH \Mbh\ using Eq.~(11) of \cite{Davis_19}:
\begin{eqnarray}
M_\bullet = 10^{7.24\pm0.82} \Msun \left(\frac{M_{\star,\mathrm{bulge}}}{1.15\times 10^{10}\Msun}\right)^{2.44\pm0.35} \, ;
\end{eqnarray}
\item  We compute the Sérsic index of the bulge $n_\mathrm{bulge}$ using Eq.~(12) of \cite{Davis_19}:
\begin{eqnarray}
n_\mathrm{bulge} = 2.20 \left(\frac{M_\bullet}{10^{7.45\pm0.84}\Msun}\right)^{1/(2.76\pm0.70)} \, ;
\end{eqnarray}
\item A random number $f$ is uniformly drawn in [0,1]. If $f~<~f_\mathrm{NSC}(M_\mathrm{gal,star})$, where $f_\mathrm{NSC}(M_\mathrm{gal,star})$ is the nuclear fraction for galaxies with mass $M_\mathrm{gal,star}$, then we place a NSC in the galaxy and go to step (vii) and further. In the other situation, no NSC is added. {$f_\mathrm{NSC}(M_\mathrm{gal,star})$ is obtained fitting Fig.~8 of \cite{SanchezJanssen_19} with:}
\begin{eqnarray}
f_\mathrm{NSC} = 0.9 \times \exp {\left[ - \left(\frac{\log{\left(\frac{M_\mathrm{gal,star}}{\Msun}\right)} - 8.8}{1.1} \right)^2 \right]};
\end{eqnarray}
\item We compute the mass of the NSC $M_{\star,\mathrm{NSC}}$ using Eq.~(6) of \cite{Pechetti_19}:
\begin{eqnarray}
M_{\star,\mathrm{NSC}} = 10^{6\pm0.13}\Msun \left(\frac{M_\mathrm{gal,star}}{10^{8.88}\Msun}\right)^{0.91} \, ;
\end{eqnarray}
\item We compute the effective radius of the NSC $R_\mathrm{eff, NSC}$ using Eq.~(4) of \cite{Dabringhausen_08} (reported in Eq.~\eqref{eq:Dabringausen});
\item We compute the Sérsic index $n_\mathrm{NSC}$ fitting Fig.~8 of \cite{Pechetti_19} with:
\begin{eqnarray}
\log{n_\mathrm{NSC}} = (-0.245 {\pm 0.094}) \log{\left(\frac{M_\mathrm{\star, NSC}}{\Msun}\right)} + (2.10 {\pm 0.93}) \, . \label{eq:nNSCofMNSC}
\end{eqnarray}
\end{enumerate}

For all steps but (ii) and (vi), the fitted parameters used in the relations are drawn from normal distributions $\mathcal{N}(\mu,\sigma)$ with mean $\mu$ and standard deviation $\sigma$ given by the different authors (the $\mu\pm\sigma$ in the above Equations). For instance, the parameter $a$ from Eq.~(4) in \cite{Dabringhausen_08} (our Eq.~\eqref{eq:Dabringausen}), used to infer the effective radius of both the bulge and the NSC, is drawn in $\mathcal{N}(2.95,0.24)$. This is done to take into account the scatter in the scaling relations.

As we assume many relations with their scatter, our method sometime produces ``irrealistic'' galaxies. In particular, the Sérsic indices could be negative or arbitrarily large: we remove galaxies for which the Sérsic index of the bulge is not in the interval $[0.5, 10]$ (reducing the sample by 2/3) and, among the remaining galaxies which have a NSC, we remove those for which the Sérsic index of the NSC is not in $[0.5, 10]$ (reducing again by 1/3). We also remove structures for which the MBH is more massive than the bulge or the NSC ($\sim1,000$ cases), resulting in a final sample of $\sim25,000$ galaxies.

For all galaxies with a MBH less massive than $10^8\Msun$, we compute the TDE rate using the technique described in \S\ref{sec:EstimateOfTheTDERate} (we can afford to use \textsc{PhaseFlow} as the number of structures remains fairly small). Similarly to \S\ref{sec:Data}, for galaxies which have a NSC, we compute the TDE rate originating from stars in the bulge and the NSC separately in order to study their respective contribution, and the total TDE rate is simply the sum of the two. The TDE rate in galaxies with MBHs more massive than $10^8\Msun$ is set to 0 to take into account that solar like stars would be swallowed whole and no TDE would be seen.

\subsection{Results}
\label{sec:Results}
\begin{figure*}
\includegraphics[width=\columnwidth]{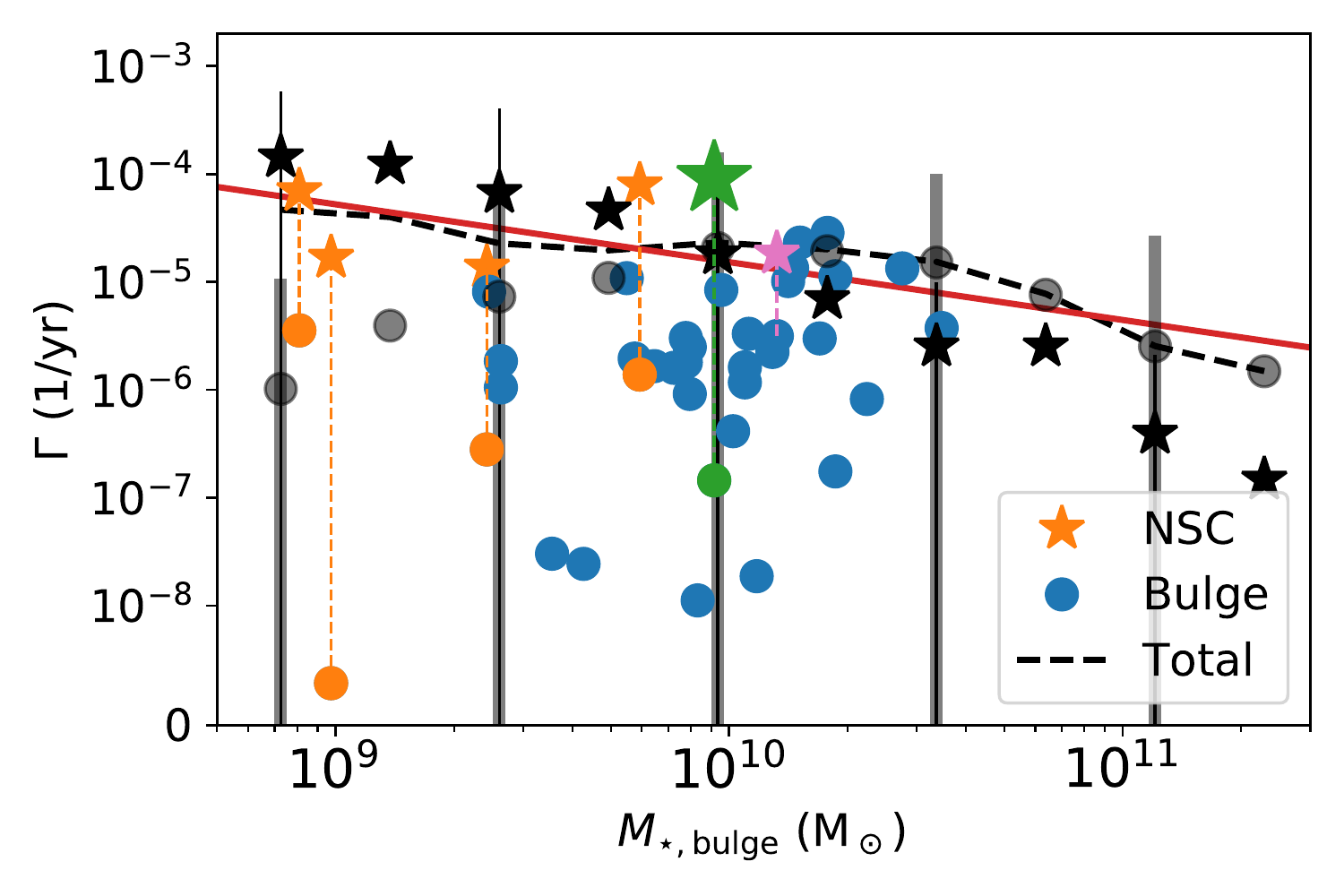} 
\hfill
\includegraphics[width=\columnwidth]{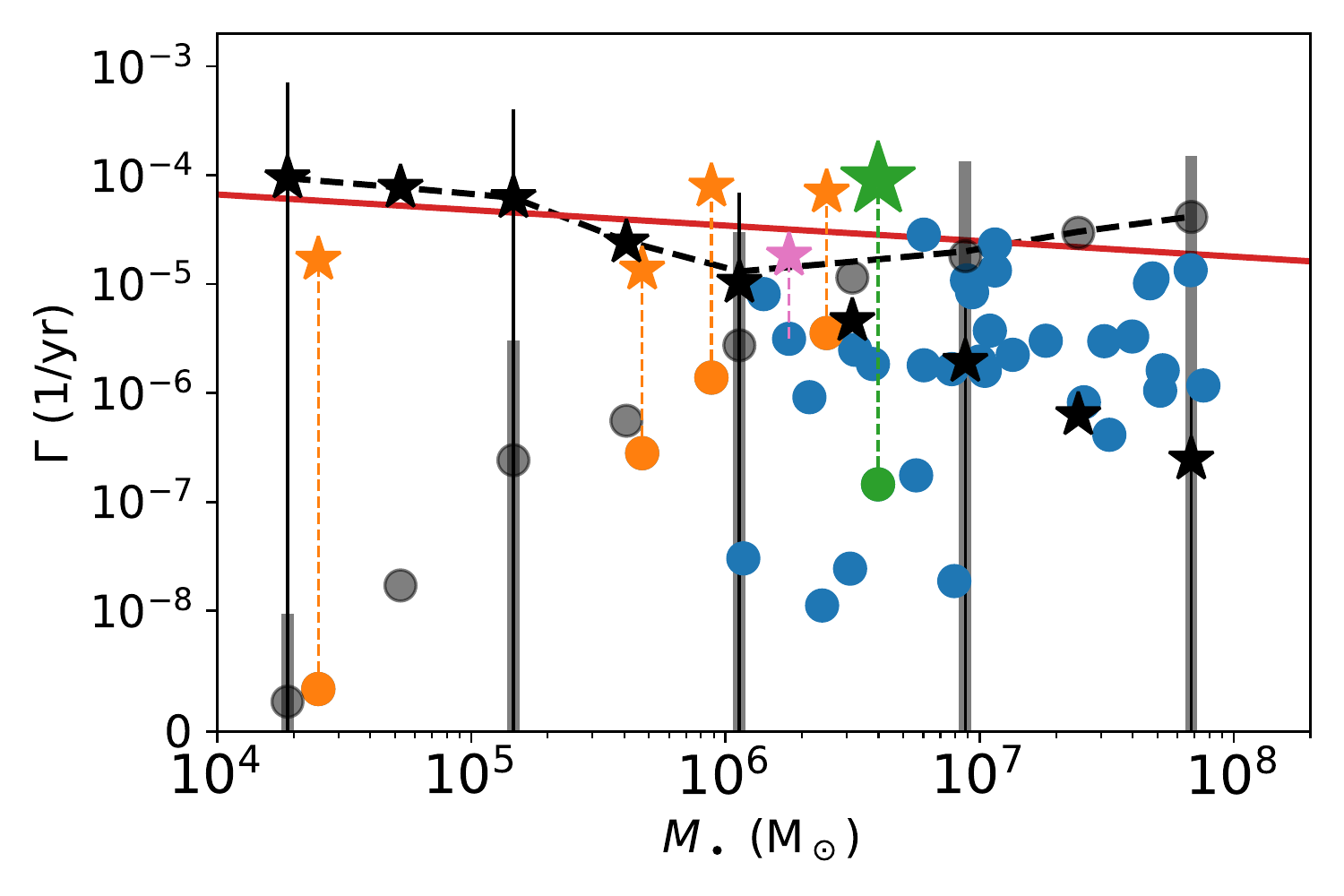} \\
\includegraphics[width=\textwidth]{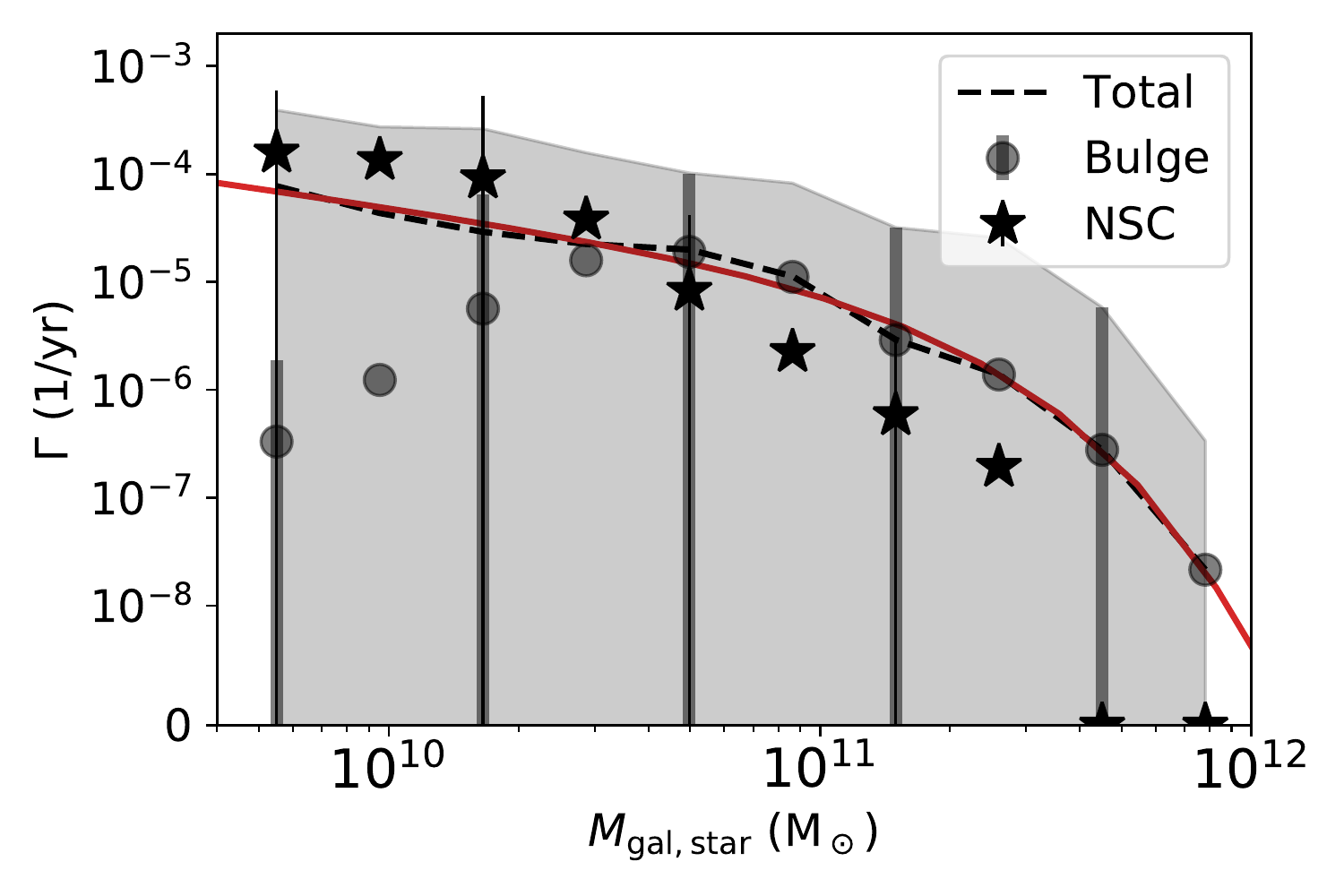} 
\caption{\textbf{Top left:} TDE rate as a function of the mass of the bulge. \textbf{Top right:} TDE rate as a function of the mass of the MBH. \textbf{Bottom:} TDE rate as a function of the mass of the galaxy. In all cases we show the mean contribution from the bulge (circles), for galaxies with a NSC we show the mean contribution from the NSC to probe their relative contribution with respect to bulges. For all galaxies we show the mean total (bulge+NSC) TDE rate with the black dashed line  and the fit in red is from Eq.~\eqref{eq:GammaOfMbulge}. Coloured markers represent real galaxies (see \S\ref{sec:Data}) while black represents the results of our model (see \ref{sec:MockCatalog}). The variance of the TDE rate originated from stars in the bulge/NSC is marked with error bars, and of the total TDE rate with a grey shaded area: a null TDE rate is within 1 sigma at all masses. Rates sourced from NSCs are typically 2/4/2 orders of magnitude larger than from bulges for light ($10^{9}\Msun$/$10^{4}\Msun$/$5\times10^{9}\Msun$) bulges/MBHs/galaxies, but their contribution is negligible for more massive objects. Overall, the total TDE rate is fairly constant at $10^{-5}\yr\mo$ and declines when galaxies are more massive than $10^{11}\Msun$ and contain MBHs swallowing stars whole and resulting in no observable TDE.}

\label{fig:GammaModel}
\end{figure*}

\begin{figure*}
\includegraphics[width=\columnwidth]{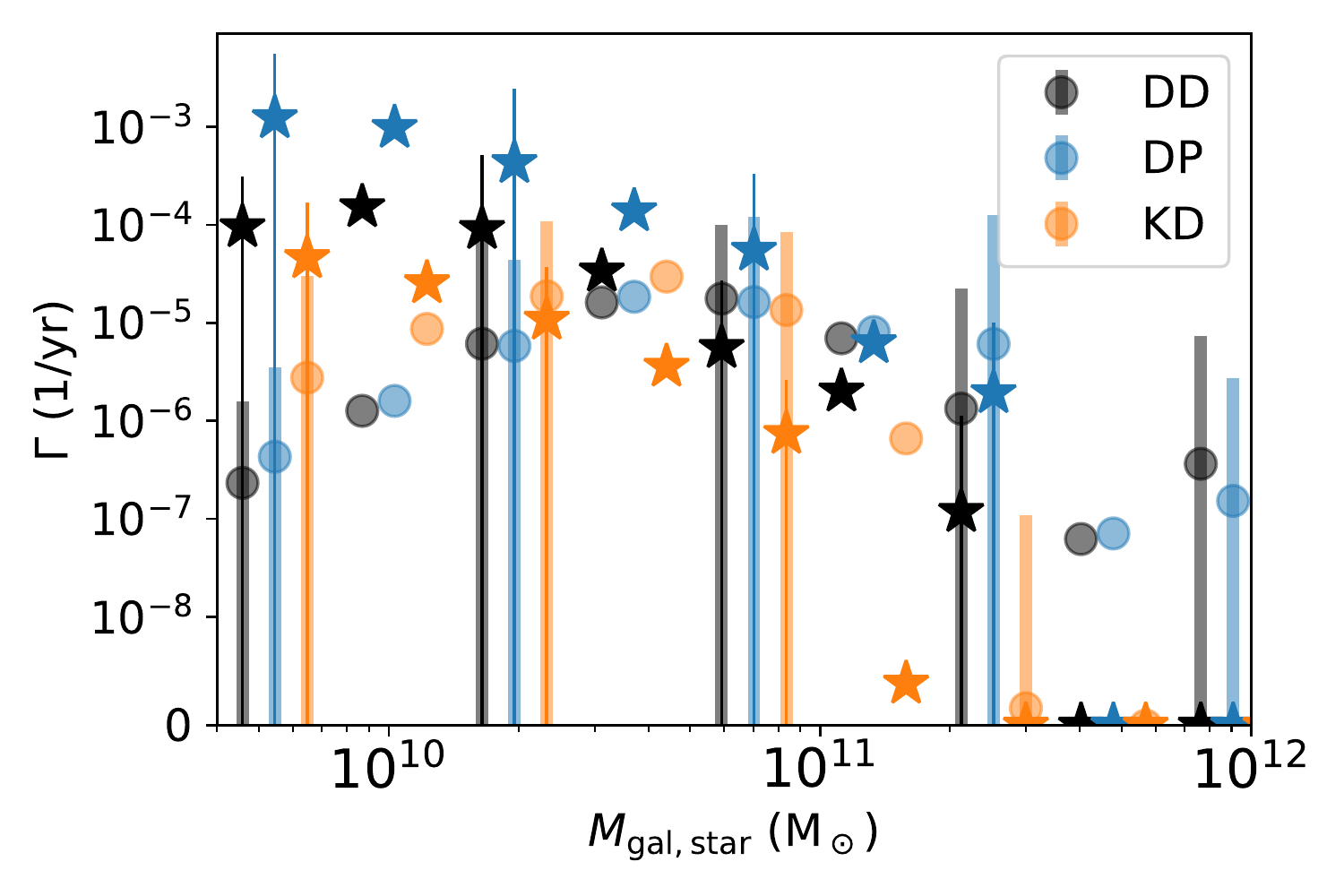} 
\hfill
\includegraphics[width=\columnwidth]{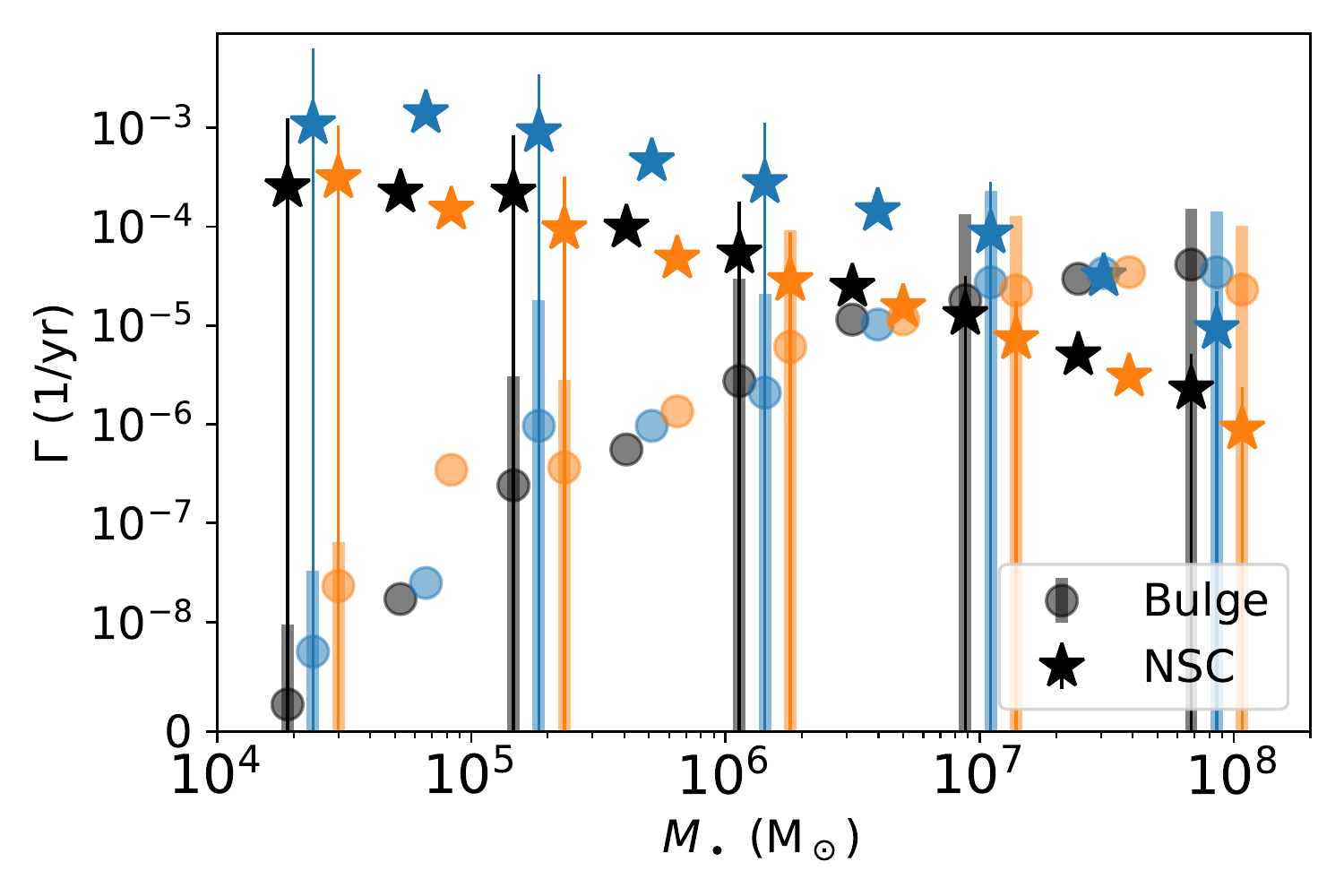} 
\caption{\textbf{Left:} TDE rate as a function of the mass of the galaxy. \textbf{Right:} TDE rate as a function of the mass of the MBH. In both cases, contribution from the bulge (circles) and from the NSC (stars) are shown for the 3 models described in Table~\ref{table:ScalingRelations}.}
\label{fig:GammaModelMany}
\end{figure*}

We show in Fig.~\ref{fig:GammaModel} the TDE rate generated from stars in the bulge (circles) and, for galaxies which have a NSC, the TDE rate originating from stars in the NSC (stars) as a function of the mass of the bulge (upper left panel), of the MBH (upper right panel) and of the galaxy (lower panel). The mean total (bulge+NSC) TDE rate including all galaxies, even those for which the TDE rate is 0, is shown with the black dashed line. The error bars simply indicate the variance at fixed mass, showing that a null TDE rate is within a 1-$\sigma$ error at all masses. For the two upper panels we also show the TDE rate of the ``real'' galaxies analyzed in \S\ref{sec:ApplicationToRealData}.

Overall, our model is in good agreement with observations, with most of the TDE rates of observed galaxies being at less than 1-$\sigma$ from the mean value of our model. Given the large size of our mock catalog, we can now investigate trends.

We start with the TDE rate as estimated when only the contribution of the bulge is included. It is somewhat similar in the three panels\footnote{We show the top two panels to overplot our estimated TDE rate from observed galaxies, and the lower panel is useful for comparison with observations for which the mass of the galaxy host is known.}: the rate increases when the mass of the bulge/MBH/galaxy increases, until it decreases at $3\times 10^{10}\Msun$/$10^{8}\Msun$/$10^{11}\Msun$, mostly because MBHs become more massive than the adopted threshold of $10^{8}\Msun$ for stars to be swallowed whole resulting in no observed TDEs.

{For galaxies which have a NSC, we compute the mean TDE rate originated from stars in NSCs. This allows to study the relative contribution of bulges and NSCs. We find that rates sourced from NSCs are typically 2/4/2 orders of magnitude larger than from bulges for light ($10^{9}\Msun$/$10^{4}\Msun$/$5\times10^{9}\Msun$) bulges/MBHs/galaxies. This confirms our expectations from our observational sample: it is necessary to resolve NSCs if one wants to properly estimate the TDE rate of light bulges/MBHs/galaxies. If one only needs an order-of-magnitude estimate, then it suffices to know if an NSC is present, since the rates are generally between $10^{-5}-10^{-4} \yr\mo$ for light MBHs hosted in NSCs.}

{Moving to more massive objects (bulge/MBH/galaxy more massive than $10^{10}\Msun$/$10^{6}\Msun$/$3\times10^{10}\Msun$), even when NSCs are present, their contribution to the TDE rate becomes smaller than that of bulges: it is not necessary to resolve or at least know if an NSC is present if one wants to estimate the TDE rate of massive bulges/MBHs/galaxies.}

When the fraction of nucleated galaxies is taken into account, we can estimate the mean total (bulge+NSC) TDE rate (dashed line). It is fairly constant with the bulge/MBH/galaxy mass and equals few $10^{-5}\yr^{-1}$ until it drops at $3\times 10^{10}\Msun$/$10^{8}\Msun$/$10^{11}\Msun$, when stars are swallowed whole and not tidally disrupted. To be more precise, we use a least-square regression on the mean total TDE rate to obtain:
\begin{eqnarray}
\frac{\Gamma}{\yr\mo} &=& 10^{-4.5\pm0.5} \left( \frac{M_\bullet}{10^6 \Msun} \right)^{-0.14\pm0.08}  \label{eq:GammaOfMBH} \nonumber\\
\frac{\Gamma}{\yr\mo} &=& 10^{-4.8\pm0.8} \left( \frac{M_\mathrm{\star,bulge}}{10^{10} \Msun} \right)^{-0.54\pm0.08} \label{eq:GammaOfMbulge} \\
\frac{\Gamma}{\yr\mo} &=& 10^{-5.1\pm0.1} \left( \frac{M_\mathrm{gal}}{10^{11} \Msun} \right)^{-0.55\pm0.08} e^{ - (3.3\pm0.3) \frac{M_\mathrm{gal}}{10^{11} \Msun} } \label{eq:GammaOfMgal} \, ,  \nonumber
\end{eqnarray}
with respective variance about the fit of $10^{-3.3}\yr\mo$, $10^{-2.3}\yr\mo$ and $10^{-3.3}\yr\mo$ (in this situation, we give the variance in linear space in order to take into account that some systems have a null TDE rate). Again, at all bulge/MBH/galaxy mass, a null TDE rate is within 1-$\sigma$.

The rates (few $10^{-5}\yr^{-1}$ per galaxy) are in agreement with current \textit{observed} TDE rates \citep{Donley_02, VanVelzen_14, Auchettl_18}. {In addition, $\Gamma\propto M^{-0.14}_\bullet$ is in agreement with \cite{VanVelzen_18} who finds that $\Gamma\propto M^0_\bullet$ is more consistent with \textit{observed} data than $\Gamma\propto M^{0.3}_\bullet$ or $\Gamma\propto M^{-0.5}_\bullet$. However, our results differ from \cite{Stone_16a} who find $\Gamma\propto M^{-0.404}_\bullet$. This is mainly because they kept MBHs with $M_\bullet>10^8\Msun$ in their sample. Such MBHs have a very low TDE rate that steepens the slope, and if we re-fit their sample (using their Table~C1), keeping only MBHs with mass $<10^8\Msun$, as we do in order to consider only observable TDEs, we obtain $\Gamma=10^{-4.2} \yr\mo (M_\bullet/10^6\Msun)^{0.02}$, similar to what we found.}

It can be somewhat surprising that we recover similar results as \cite{Stone_16a} and \cite{VanVelzen_18} who do \textit{not} consider that galaxies host NSCs, which, as we have shown, can significantly enhance the TDE rate. The reasons for this are twofold. Firstly, \cite{Stone_16a} and \cite{VanVelzen_18} considered MBHs more massive than $\sim10^6\Msun$, where the contribution of NSCs is actually negligible. Secondly, we considered that the different components of galaxies (bulge and NSC) can be well approximated with a Sérsic profile, while \cite{Stone_16a} used a Nuker profile. Discussion on the goodness of these profiles is beyond the scope of this paper, but the deprojected Sérsic profile is a Prugniel profile which has a fairly flat inner 3D logarithmic slope ($p<1$), contrary to the deprojected Nuker profile which can have steeper inner 3D logarithmic slope, resulting in larger TDE rates. In terms of density profile modelling, therefore, our results should be considered lower limits to the expected TDE rates.

{We predict that the TDE rate remains constant with the mass of the MBH down to masses of $10^4\Msun$. To date, the observed rate below $10^6\Msun$ is essentially unconstrained. It is true that the \textit{observed} number of TDEs drops below $M_\bullet \sim 10^6\Msun$ \citep{Stone_16a, Wevers_19}, but we recall that this could simply be that these TDEs are extremely challenging to observe due to their faint luminosity \citep{Guillochon_13, Piran_15, Dai_15, Roth_16, Dai_18,  Mockler_19}.}

An important point is that we have also assumed that all galaxies harbour a MBH, and while this is probably a good assumption for massive galaxies, it is not the case in the low mass regime, where the occupation fraction is theoretically predicted to decrease \citep{Volonteri_08}. \cite{Stone_16a} explore the effect of taking into account a MBH occupation fraction that depends on the bulge mass, assuming a one-to-one relation with MBH mass. In this paper we do not include this additional parameter, since its functional form is very uncertain, both theoretically and observationally \citep{Greene_19}. A drop in the TDE rate below our predictions at low galaxy mass would be a hint that that the MBH occupation fraction in dwarfs is not 100\%. 

While the effect of NSCs is negligible in massive ($>3\times 10^{10}\Msun$) galaxies, they form a particularly important component in the dwarf regime \citep[we recall that 90\% of $10^9\Msun$ galaxies, and more than 50\% of $10^8\Msun$ galaxies, have a NSC;][]{SanchezJanssen_19}, enhancing the TDE rate by few orders of magnitude. We note that \cite{Biava_19}, who studied the evolution of lifetime of MBH binaries in the context of gravitational waves for LISA, also find that estimates of the lifetimes of the most massive binaries (in massive galaxies) is not strongly dependent on the details of the central density profile.
However, the low mass binary regime is strongly affected by details of the stellar density profile and the presence, or not, of a NSC, with binary lifetimes varying in between 10~Myr, in cases with NSCs, to 100~Gyr in cases without NSCs for $10^5\Msun$ binaries. This suggests that, with the hundreds to thousands of TDEs which will be detected with the LSST \citep{VanVelzen_11} or eROSITA \citep{Jonker_20}, we will learn about the internal structure of dwarf galaxies, which will be useful in making predictions for gravitational wave detection with LISA.

\section{Effects of uncertainties in the scaling relations}
\label{sec:EffectsOfTheScalingRelations}

\begin{table}
\begin{center}
\begin{tabular}{p{1.5cm}p{2.7cm}p{3.1cm}|}
\hline
\textbf{Name} & $M_\bullet - M_\mathrm{\star,bulge}$ & $M_{\star\mathrm{, NSC}} - R_\mathrm{eff, NSC}$\\
\hline
\hline
DD (fiducial)  & \cite{Davis_19} & \cite{Dabringhausen_08} \\
DP & \cite{Davis_19} & \cite{Pechetti_19} \\
KD  & \cite{Kormendy_13} & \cite{Davis_19} \\
\hline
\end{tabular}
\caption{Name of the models and scaling relations used. Our fiducial model is the one described in \S\ref{sec:MockCatalog}}
\label{table:ScalingRelations}
\end{center}
\end{table}

In order to explore trends with MBH, bulge and galaxy mass, we built a mock catalog using a set of scaling relations. However, the physical meaning of these relations is still debated, and different groups, using different samples, find different relations. While we partly took this into account by including scatter about the relations (see \S\ref{sec:MockCatalog}), we adopt here another approach, using different sets of scaling relations. In particular, we re-perform the exact same procedure as in \S\ref{sec:MockCatalog} (see Table~\ref{table:ScalingRelations} for the different cases studied) but we use the $M_\bullet -  M_\mathrm{\star,bulge}$ relation from \cite{Kormendy_13}:
\begin{eqnarray}
M_\bullet =(0.49\pm0.06)\times 10^{9\pm0.28}\Msun\left( \frac{M_\mathrm{\star, bulge}}{10^{11}\Msun}\right)^{1.17\pm0.08}
\end{eqnarray}
or the $M_{\star\mathrm{, NSC}} - R_\mathrm{eff, NSC}$ relation from  \cite{Pechetti_19}:
\begin{eqnarray}
R_\mathrm{eff,NSC}=10^{-1.605\pm0.06}\pc \left( \frac{M_\mathrm{\star, NSC}}{\Msun} \right)^{0.333} \, .
\end{eqnarray}

We show in Fig.~\ref{fig:GammaModelMany} the TDE rate sourced from the NSC (stars) and from the bulge (circles), for these three models, as a function of the mass of the galaxy (left panel) and of the MBH (right panel). {Using the $M_\bullet -  M_\mathrm{\star,bulge}$ relation from \cite{Davis_19} (black) or \cite{Kormendy_13} (orange) gives similar results because the scatter in these relations is so large \citep[$\pm1\dex$ in ][]{Davis_19} that the details in the relation do not impact the mean TDE rate. On the other hand, when we use the $M_{\star\mathrm{, NSC}}~-~R_\mathrm{eff, NSC}$ relation from \cite{Pechetti_19} (blue) instead of \cite{Dabringhausen_08} (black), the final rates differ by $\sim1\dex$. The reason is that \cite{Pechetti_19} predict effective radii for NSCs that are 2-8 times smaller than \cite{Dabringhausen_08}, and from Eqs.~\eqref{eq:rho0},~\eqref{eq:r_inf} and \eqref{eq:Gammaofrhoinf}  $\Gamma\propto R_\mathrm{eff}^{-3\times0.65}\propto R_\mathrm{eff}^\mathrm{-1.95}$, naturally resulting in rates that are 5-50 times larger with the relation of \cite{Pechetti_19}.}

Overall, the details vary from one model to the other, but they  always remain within the scatter so that our conclusions are unaffected.

\section{Conclusions}
\label{sec:Conclusions}
We have estimated the TDE rate in 37 galaxies for which we have the stellar surface density profile, a dynamical estimate of the mass of the MBH, and 6 of which, including our Milky Way, have a resolved NSC. We also estimated the TDE rate  in a mock catalog of 25,000 galaxies  built using a set of scaling relations, including the nucleated fraction of galaxies. Our main findings are the following:
\begin{itemize}
\item It is necessary to resolve the central part of dwarf  galaxies with masses lower than $3\times 10^{10}\Msun$ to properly estimate the TDE rate around MBHs with masses lower than $10^6\Msun$. Indeed, these galaxies may harbour a NSC, possibly enhancing the total TDE rate by 1-2 orders of magnitude. 
\item Since we assumed an occupation fraction of 100\%, a lower TDE rate in dwarfs could be a hint that this fraction is in fact lower in this regime \citep{Stone_16a}, but more work is needed to better understand whether such TDEs can be effectively discovered using current surveys.
\item The TDE rate in the Milky Way around Sagittarius A$^*$ is predicted to be $9.1~\times~10^{-5} \yr^{-1}$.
\item The TDE rate is roughly constant at few $10^{-5}\yr^{-1}$ for bulges/MBHs/galaxies up to $3\times 10^{10}\Msun$/$10^{8}\Msun$/$10^{11}\Msun$, after which stars are swallowed whole and not tidally disrupted, resulting in no observed TDEs. This result is independent of the scaling relations used, however, at fixed bulge/MBH/galaxy mass, the scatter in the TDE rate is large enough so that a null TDE rate is always possible. 
\item We provide fitting formulae giving the mean TDE rate as a function of MBH/bulge/galaxy mass (Eq.~\eqref{eq:GammaOfMbulge}).
\item If the mass of the MBH and its surrounding stellar density profile are known, one can rapidly estimate the TDE rate using the density at the influence radius (Eq.~\eqref{eq:Gammaofrhoinf}).
\end{itemize}

{We stress here again that our estimates of the TDE rate is based on the loss-cone formalism, which does not include the ``physics'' of TDEs, therefore they are upper limits to the \textit{observable} TDE rate. Nonetheless, we have shown that the TDE rates depend sensitively on the inner structures of the host galaxies on pc scales. In addition, not just the TDE rate, but also the merger rate of MBH binaries detectable as gravitational wave sources depends on the stellar distribution near MBHs and the presence (or absence) of a NSC \citep{Biava_19}. In summary, with a better understanding of the physics relevant for TDE flare emissions, the observed TDE rate and luminosity function can be used to fill the gap in constraining the stellar density, slope and structures in the vicinity of MBHs, especially for dwarf galaxies. With future observations of TDEs with the LSST or eROSITA, which will more precisely constrain the TDE rate, we will refine our predictions for MBH binary hardening rates and therefore MBH merger rates for LISA. This comparison is subject of an ongoing study.}

\section*{Acknowledgments}

{We thank the anonymous referee for providing comments that greatly improved the quality of the paper.} HP and JD are indebted to the Danish National Research Foundation (DNRF132) and the Hong Kong government (GRF grant HKU27305119) for support. MV thanks Rainer Sch\"odel for help with models of the NSC of the Milky Way. The authors thank the Yukawa Institute for Theoretical Physics at Kyoto University. Discussions during the YITP workshop YITP-T-19-07 on International Molecule-type Workshop "Tidal Disruption Events: General Relativistic Transients" were useful to complete this work. MC acknowledges funding from MIUR under the grant PRIN 2017-MB8AEZ.\\

\section*{Data availability}
Scripts and data used in this paper are available upon request.

\bsp	\label{lastpage}

\begin{thebibliography}{}
\makeatletter
\relax
\def\mn@urlcharsother{\let\do\@makeother \do\$\do\&\do\#\do\^\do\_\do\%\do\~}
\def\mn@doi{\begingroup\mn@urlcharsother \@ifnextchar [ {\mn@doi@}
  {\mn@doi@[]}}
\def\mn@doi@[#1]#2{\def\@tempa{#1}\ifx\@tempa\@empty \href
  {http://dx.doi.org/#2} {doi:#2}\else \href {http://dx.doi.org/#2} {#1}\fi
  \endgroup}
\def\mn@eprint#1#2{\mn@eprint@#1:#2::\@nil}
\def\mn@eprint@arXiv#1{\href {http://arxiv.org/abs/#1} {{\tt arXiv:#1}}}
\def\mn@eprint@dblp#1{\href {http://dblp.uni-trier.de/rec/bibtex/#1.xml}
  {dblp:#1}}
\def\mn@eprint@#1:#2:#3:#4\@nil{\def\@tempa {#1}\def\@tempb {#2}\def\@tempc
  {#3}\ifx \@tempc \@empty \let \@tempc \@tempb \let \@tempb \@tempa \fi \ifx
  \@tempb \@empty \def\@tempb {arXiv}\fi \@ifundefined
  {mn@eprint@\@tempb}{\@tempb:\@tempc}{\expandafter \expandafter \csname
  mn@eprint@\@tempb\endcsname \expandafter{\@tempc}}}

\bibitem[\protect\citeauthoryear{{Aharon}, {Mastrobuono Battisti}  \&
  {Perets}}{{Aharon} et~al.}{2016}]{Aharon_16}
{Aharon} D.,  {Mastrobuono Battisti} A.,   {Perets} H.~B.,  2016, \mn@doi [ApJ]
  {10.3847/0004-637X/823/2/137}, \href
  {https://ui.adsabs.harvard.edu/abs/2016ApJ...823..137A} {823, 137}

\bibitem[\protect\citeauthoryear{{Alexander} \& {Bar-Or}}{{Alexander} \&
  {Bar-Or}}{2017}]{Alexander_17}
{Alexander} T.,  {Bar-Or} B.,  2017, \mn@doi [Nature Astronomy]
  {10.1038/s41550-017-0147}, \href
  {http://adsabs.harvard.edu/abs/2017NatAs...1E.147A} {1, 0147}

\bibitem[\protect\citeauthoryear{{Amaro-Seoane} et~al.,}{{Amaro-Seoane}
  et~al.}{2017}]{LISA2017}
{Amaro-Seoane} P.,  et~al., 2017, arXiv e-prints, \href
  {https://ui.adsabs.harvard.edu/abs/2017arXiv170200786A} {p. arXiv:1702.00786}

\bibitem[\protect\citeauthoryear{{Arca-Sedda} \&
  {Capuzzo-Dolcetta}}{{Arca-Sedda} \& {Capuzzo-Dolcetta}}{2017}]{ArcaSedda_17}
{Arca-Sedda} M.,  {Capuzzo-Dolcetta} R.,  2017, \mn@doi [MNRAS]
  {10.1093/mnras/stx1586}, \href
  {https://ui.adsabs.harvard.edu/abs/2017MNRAS.471..478A} {471, 478}

\bibitem[\protect\citeauthoryear{{Auchettl}, {Guillochon}  \&
  {Ramirez-Ruiz}}{{Auchettl} et~al.}{2017}]{Auchettl_17}
{Auchettl} K.,  {Guillochon} J.,   {Ramirez-Ruiz} E.,  2017, \mn@doi [ApJ]
  {10.3847/1538-4357/aa633b}, \href
  {https://ui.adsabs.harvard.edu/abs/2017ApJ...838..149A} {838, 149}

\bibitem[\protect\citeauthoryear{{Auchettl}, {Ramirez-Ruiz}  \&
  {Guillochon}}{{Auchettl} et~al.}{2018}]{Auchettl_18}
{Auchettl} K.,  {Ramirez-Ruiz} E.,   {Guillochon} J.,  2018, \mn@doi [ApJ]
  {10.3847/1538-4357/aa9b7c}, \href
  {http://adsabs.harvard.edu/abs/2018ApJ...852...37A} {852, 37}

\bibitem[\protect\citeauthoryear{{Biava}, {Colpi}, {Capelo}, {Bonetti},
  {Volonteri}, {Tamfal}, {Mayer}  \& {Sesana}}{{Biava} et~al.}{2019}]{Biava_19}
{Biava} N.,  {Colpi} M.,  {Capelo} P.~R.,  {Bonetti} M.,  {Volonteri} M.,
  {Tamfal} T.,  {Mayer} L.,   {Sesana} A.,  2019, \mn@doi [MNRAS]
  {10.1093/mnras/stz1614}, \href
  {https://ui.adsabs.harvard.edu/abs/2019MNRAS.487.4985B} {487, 4985}

\bibitem[\protect\citeauthoryear{Binney \& Tremaine}{Binney \&
  Tremaine}{1987}]{BT_87}
Binney J.,  Tremaine S.,  1987, Galactic Dynamics, first edn.
Princeton Series in Astrophysics, Princeton University Press

\bibitem[\protect\citeauthoryear{{Bullock} \& {Boylan-Kolchin}}{{Bullock} \&
  {Boylan-Kolchin}}{2017}]{Bullock_17}
{Bullock} J.~S.,  {Boylan-Kolchin} M.,  2017, \mn@doi [ARAA]
  {10.1146/annurev-astro-091916-055313}, \href
  {https://ui.adsabs.harvard.edu/abs/2017ARA&A..55..343B} {55, 343}

\bibitem[\protect\citeauthoryear{{Dabringhausen}, {Hilker}  \&
  {Kroupa}}{{Dabringhausen} et~al.}{2008}]{Dabringhausen_08}
{Dabringhausen} J.,  {Hilker} M.,   {Kroupa} P.,  2008, \mn@doi [MNRAS]
  {10.1111/j.1365-2966.2008.13065.x}, \href
  {https://ui.adsabs.harvard.edu/abs/2008MNRAS.386..864D} {386, 864}

\bibitem[\protect\citeauthoryear{{Dai}, {McKinney}  \& {Miller}}{{Dai}
  et~al.}{2015}]{Dai_15}
{Dai} L.,  {McKinney} J.~C.,   {Miller} M.~C.,  2015, \mn@doi [ApJ]
  {10.1088/2041-8205/812/2/L39}, \href
  {https://ui.adsabs.harvard.edu/abs/2015ApJ...812L..39D} {812, L39}

\bibitem[\protect\citeauthoryear{{Dai}, {McKinney}, {Roth}, {Ramirez-Ruiz}  \&
  {Miller}}{{Dai} et~al.}{2018}]{Dai_18}
{Dai} L.,  {McKinney} J.~C.,  {Roth} N.,  {Ramirez-Ruiz} E.,   {Miller} M.~C.,
  2018, \mn@doi [ApJ] {10.3847/2041-8213/aab429}, \href
  {http://adsabs.harvard.edu/abs/2018ApJ...859L..20D} {859, L20}

\bibitem[\protect\citeauthoryear{{Davis}, {Graham}  \& {Cameron}}{{Davis}
  et~al.}{2019}]{Davis_19}
{Davis} B.~L.,  {Graham} A.~W.,   {Cameron} E.,  2019, \mn@doi [ApJ]
  {10.3847/1538-4357/aaf3b8}, \href
  {https://ui.adsabs.harvard.edu/abs/2019ApJ...873...85D} {873, 85}

\bibitem[\protect\citeauthoryear{{Dong} et~al.,}{{Dong} et~al.}{2016}]{Dong_16}
{Dong} S.,  et~al., 2016, \mn@doi [Science] {10.1126/science.aac9613}, \href
  {https://ui.adsabs.harvard.edu/abs/2016Sci...351..257D} {351, 257}

\bibitem[\protect\citeauthoryear{Donley, Brandt, Eracleous  \& Boller}{Donley
  et~al.}{2002}]{Donley_02}
Donley J.~L.,  Brandt W.~N.,  Eracleous M.,   Boller T.,  2002, \mn@doi [The
  Astronomical Journal] {10.1086/342280}, 124, 1308

\bibitem[\protect\citeauthoryear{{Dubois}, {Volonteri}, {Silk}, {Devriendt},
  {Slyz}  \& {Teyssier}}{{Dubois} et~al.}{2015}]{Dubois_15}
{Dubois} Y.,  {Volonteri} M.,  {Silk} J.,  {Devriendt} J.,  {Slyz} A.,
  {Teyssier} R.,  2015, \mn@doi [MNRAS] {10.1093/mnras/stv1416}, \href
  {https://ui.adsabs.harvard.edu/abs/2015MNRAS.452.1502D} {452, 1502}

\bibitem[\protect\citeauthoryear{{French}, {Arcavi}  \& {Zabludoff}}{{French}
  et~al.}{2016}]{French_16}
{French} K.~D.,  {Arcavi} I.,   {Zabludoff} A.,  2016, \mn@doi [ApJ]
  {10.3847/2041-8205/818/1/L21}, \href
  {http://adsabs.harvard.edu/abs/2016ApJ...818L..21F} {818, L21}

\bibitem[\protect\citeauthoryear{{French}, {Wevers}, {Law-Smith}, {Graur}  \&
  {Zabludoff}}{{French} et~al.}{2020}]{French_20b}
{French} K.~D.,  {Wevers} T.,  {Law-Smith} J.,  {Graur} O.,   {Zabludoff}
  A.~I.,  2020, arXiv e-prints, \href
  {https://ui.adsabs.harvard.edu/abs/2020arXiv200302863F} {p. arXiv:2003.02863}

\bibitem[\protect\citeauthoryear{{Gezari} et~al.,}{{Gezari}
  et~al.}{2012}]{Gezari_12}
{Gezari} S.,  et~al., 2012, \mn@doi [Nature] {10.1038/nature10990}, \href
  {http://adsabs.harvard.edu/abs/2012Natur.485..217G} {485, 217}

\bibitem[\protect\citeauthoryear{{Graur}, {French}, {Zahid}, {Guillochon},
  {Mandel}, {Auchettl}  \& {Zabludoff}}{{Graur} et~al.}{2018}]{Graur_18}
{Graur} O.,  {French} K.~D.,  {Zahid} H.~J.,  {Guillochon} J.,  {Mandel} K.~S.,
   {Auchettl} K.,   {Zabludoff} A.~I.,  2018, \mn@doi [ApJ]
  {10.3847/1538-4357/aaa3fd}, \href
  {http://adsabs.harvard.edu/abs/2018ApJ...853...39G} {853, 39}

\bibitem[\protect\citeauthoryear{{Greene}, {Strader}  \& {Ho}}{{Greene}
  et~al.}{2019}]{Greene_19}
{Greene} J.~E.,  {Strader} J.,   {Ho} L.~C.,  2019, arXiv e-prints, \href
  {https://ui.adsabs.harvard.edu/abs/2019arXiv191109678G} {p. arXiv:1911.09678}

\bibitem[\protect\citeauthoryear{{Guillochon} \& {Ramirez-Ruiz}}{{Guillochon}
  \& {Ramirez-Ruiz}}{2013}]{Guillochon_13}
{Guillochon} J.,  {Ramirez-Ruiz} E.,  2013, \mn@doi [ApJ]
  {10.1088/0004-637X/767/1/25}, \href
  {http://adsabs.harvard.edu/abs/2013ApJ...767...25G} {767, 25}

\bibitem[\protect\citeauthoryear{{Hills}}{{Hills}}{1975}]{Hills_75}
{Hills} J.~G.,  1975, \mn@doi [Nature] {10.1038/254295a0}, \href
  {http://adsabs.harvard.edu/abs/1975Natur.254..295H} {254, 295}

\bibitem[\protect\citeauthoryear{{Hung} et~al.,}{{Hung}
  et~al.}{2018}]{Hung2018}
{Hung} T.,  et~al., 2018, \mn@doi [\apjs] {10.3847/1538-4365/aad8b1}, \href
  {https://ui.adsabs.harvard.edu/abs/2018ApJS..238...15H} {238, 15}

\bibitem[\protect\citeauthoryear{{Ivanov} \& {Chernyakova}}{{Ivanov} \&
  {Chernyakova}}{2006}]{Ivanov2006}
{Ivanov} P.~B.,  {Chernyakova} M.~A.,  2006, \mn@doi [\aap]
  {10.1051/0004-6361:20053409}, \href
  {https://ui.adsabs.harvard.edu/abs/2006A&A...448..843I} {448, 843}

\bibitem[\protect\citeauthoryear{{Jonker}, {Stone}, {Generozov}, {Velzen}  \&
  {Metzger}}{{Jonker} et~al.}{2020}]{Jonker_20}
{Jonker} P.~G.,  {Stone} N.~C.,  {Generozov} A.,  {Velzen} S.~v.,   {Metzger}
  B.,  2020, \mn@doi [ApJ] {10.3847/1538-4357/ab659c}, \href
  {https://ui.adsabs.harvard.edu/abs/2020ApJ...889..166J} {889, 166}

\bibitem[\protect\citeauthoryear{{Kesden}}{{Kesden}}{2012a}]{Kesden_12}
{Kesden} M.,  2012a, \mn@doi [Physical Review] {10.1103/PhysRevD.85.024037},
  \href {https://ui.adsabs.harvard.edu/abs/2012PhRvD..85b4037K} {85, 024037}

\bibitem[\protect\citeauthoryear{{Kesden}}{{Kesden}}{2012b}]{Kesden2012}
{Kesden} M.,  2012b, \mn@doi [\prd] {10.1103/PhysRevD.85.024037}, \href
  {https://ui.adsabs.harvard.edu/abs/2012PhRvD..85b4037K} {85, 024037}

\bibitem[\protect\citeauthoryear{{Khochfar} et~al.,}{{Khochfar}
  et~al.}{2011}]{Kochfar_11}
{Khochfar} S.,  et~al., 2011, \mn@doi [MNRAS]
  {10.1111/j.1365-2966.2011.19486.x}, \href
  {https://ui.adsabs.harvard.edu/abs/2011MNRAS.417..845K} {417, 845}

\bibitem[\protect\citeauthoryear{{Kormendy} \& {Ho}}{{Kormendy} \&
  {Ho}}{2013}]{Kormendy_13}
{Kormendy} J.,  {Ho} L.~C.,  2013, \mn@doi [ARAA]
  {10.1146/annurev-astro-082708-101811}, \href
  {http://adsabs.harvard.edu/abs/2013ARA%26A..51..511K} {51, 511}

\bibitem[\protect\citeauthoryear{{Kroupa}}{{Kroupa}}{2001}]{Kroupa_01}
{Kroupa} P.,  2001, \mn@doi [MNRAS] {10.1046/j.1365-8711.2001.04022.x}, \href
  {https://ui.adsabs.harvard.edu/abs/2001MNRAS.322..231K} {322, 231}

\bibitem[\protect\citeauthoryear{{Kr{\"u}hler} et~al.,}{{Kr{\"u}hler}
  et~al.}{2018}]{Kruhler_18}
{Kr{\"u}hler} T.,  et~al., 2018, \mn@doi [AAP] {10.1051/0004-6361/201731773},
  \href {https://ui.adsabs.harvard.edu/abs/2018A&A...610A..14K} {610, A14}

\bibitem[\protect\citeauthoryear{{Lauer}, {Faber}, {Ajhar}, {Grillmair}  \&
  {Scowen}}{{Lauer} et~al.}{1998}]{Lauer_98}
{Lauer} T.~R.,  {Faber} S.~M.,  {Ajhar} E.~A.,  {Grillmair} C.~J.,   {Scowen}
  P.~A.,  1998, \mn@doi [ApJ] {10.1086/300617}, \href
  {https://ui.adsabs.harvard.edu/abs/1998AJ....116.2263L} {116, 2263}

\bibitem[\protect\citeauthoryear{{Lauer} et~al.,}{{Lauer}
  et~al.}{2007}]{Lauer_07}
{Lauer} T.~R.,  et~al., 2007, \mn@doi [ApJ] {10.1086/519229}, \href
  {http://adsabs.harvard.edu/abs/2007ApJ...664..226L} {664, 226}

\bibitem[\protect\citeauthoryear{{Law-Smith}, {Ramirez-Ruiz}, {Ellison}  \&
  {Foley}}{{Law-Smith} et~al.}{2017}]{LawSmith_17}
{Law-Smith} J.,  {Ramirez-Ruiz} E.,  {Ellison} S.~L.,   {Foley} R.~J.,  2017,
  \mn@doi [ApJ] {10.3847/1538-4357/aa94c7}, \href
  {http://adsabs.harvard.edu/abs/2017ApJ...850...22L} {850, 22}

\bibitem[\protect\citeauthoryear{{Lightman} \& {Shapiro}}{{Lightman} \&
  {Shapiro}}{1977}]{Lightman_77}
{Lightman} A.~P.,  {Shapiro} S.~L.,  1977, \mn@doi [ApJ] {10.1086/154925},
  \href {https://ui.adsabs.harvard.edu/#abs/1977ApJ...211..244L} {211, 244}

\bibitem[\protect\citeauthoryear{{Mainetti}, {Lupi}, {Campana}, {Colpi},
  {Coughlin}, {Guillochon}  \& {Ramirez-Ruiz}}{{Mainetti}
  et~al.}{2017}]{Mainetti2017}
{Mainetti} D.,  {Lupi} A.,  {Campana} S.,  {Colpi} M.,  {Coughlin} E.~R.,
  {Guillochon} J.,   {Ramirez-Ruiz} E.,  2017, \mn@doi [\aap]
  {10.1051/0004-6361/201630092}, \href
  {https://ui.adsabs.harvard.edu/abs/2017A&A...600A.124M} {600, A124}

\bibitem[\protect\citeauthoryear{{M{\'a}rquez}, {Lima Neto}, {Capelato},
  {Durret}  \& {Gerbal}}{{M{\'a}rquez} et~al.}{2000}]{Marquez_00}
{M{\'a}rquez} I.,  {Lima Neto} G.~B.,  {Capelato} H.,  {Durret} F.,   {Gerbal}
  D.,  2000, AAP, \href {https://ui.adsabs.harvard.edu/abs/2000A&A...353..873M}
  {353, 873}

\bibitem[\protect\citeauthoryear{{Mastrobuono-Battisti}, {Perets}  \&
  {Loeb}}{{Mastrobuono-Battisti} et~al.}{2014}]{MastrobuonoBattisti_14}
{Mastrobuono-Battisti} A.,  {Perets} H.~B.,   {Loeb} A.,  2014, \mn@doi [ApJ]
  {10.1088/0004-637X/796/1/40}, \href
  {https://ui.adsabs.harvard.edu/abs/2014ApJ...796...40M} {796, 40}

\bibitem[\protect\citeauthoryear{{Merritt}}{{Merritt}}{2013}]{Merrit_book}
{Merritt} D.,  2013, Dynamics and Evolution of Galactic Nuclei.
Princeton University Press

\bibitem[\protect\citeauthoryear{{Mockler}, {Guillochon}  \&
  {Ramirez-Ruiz}}{{Mockler} et~al.}{2019}]{Mockler_19}
{Mockler} B.,  {Guillochon} J.,   {Ramirez-Ruiz} E.,  2019, \mn@doi [ApJ]
  {10.3847/1538-4357/ab010f}, \href
  {https://ui.adsabs.harvard.edu/abs/2019ApJ...872..151M} {872, 151}

\bibitem[\protect\citeauthoryear{{Mummery} \& {Balbus}}{{Mummery} \&
  {Balbus}}{2020}]{Mummery_20}
{Mummery} A.,  {Balbus} S.~A.,  2020, \mn@doi [MNRAS] {10.1093/mnrasl/slaa105},
  \href {https://ui.adsabs.harvard.edu/abs/2020MNRAS.497L..13M} {497, L13}

\bibitem[\protect\citeauthoryear{{Nguyen} et~al.,}{{Nguyen}
  et~al.}{2018}]{Nguyen_18}
{Nguyen} D.~D.,  et~al., 2018, \mn@doi [ApJ] {10.3847/1538-4357/aabe28}, \href
  {http://adsabs.harvard.edu/abs/2018ApJ...858..118N} {858, 118}

\bibitem[\protect\citeauthoryear{{Pechetti}, {Seth}, {Neumayer}, {Georgiev},
  {Kacharov}  \& {den Brok}}{{Pechetti} et~al.}{2019}]{Pechetti_19}
{Pechetti} R.,  {Seth} A.,  {Neumayer} N.,  {Georgiev} I.,  {Kacharov} N.,
  {den Brok} M.,  2019, arXiv e-prints, \href
  {https://ui.adsabs.harvard.edu/abs/2019arXiv191109686P} {p. arXiv:1911.09686}

\bibitem[\protect\citeauthoryear{{Pfister}, {Bar-Or}, {Volonteri}, {Dubois}  \&
  {Capelo}}{{Pfister} et~al.}{2019}]{Pfister_19b}
{Pfister} H.,  {Bar-Or} B.,  {Volonteri} M.,  {Dubois} Y.,   {Capelo} P.~R.,
  2019, \mn@doi [MNRAS] {10.1093/mnrasl/slz091}, \href
  {https://ui.adsabs.harvard.edu/abs/2019MNRAS.tmpL..87P} {p.~L87}

\bibitem[\protect\citeauthoryear{{Piran}, {Svirski}, {Krolik}, {Cheng}  \&
  {Shiokawa}}{{Piran} et~al.}{2015}]{Piran_15}
{Piran} T.,  {Svirski} G.,  {Krolik} J.,  {Cheng} R.~M.,   {Shiokawa} H.,
  2015, \mn@doi [ApJ] {10.1088/0004-637X/806/2/164}, \href
  {https://ui.adsabs.harvard.edu/abs/2015ApJ...806..164P} {806, 164}

\bibitem[\protect\citeauthoryear{{Prugniel} \& {Simien}}{{Prugniel} \&
  {Simien}}{1997}]{Prugniel_97}
{Prugniel} P.,  {Simien} F.,  1997, AAP, \href
  {https://ui.adsabs.harvard.edu/abs/1997A&A...321..111P} {321, 111}

\bibitem[\protect\citeauthoryear{{Rees}}{{Rees}}{1988}]{Rees_88}
{Rees} M.~J.,  1988, \mn@doi [Nature] {10.1038/333523a0}, \href
  {http://adsabs.harvard.edu/abs/1988Natur.333..523R} {333, 523}

\bibitem[\protect\citeauthoryear{{Roth}, {Kasen}, {Guillochon}  \&
  {Ramirez-Ruiz}}{{Roth} et~al.}{2016}]{Roth_16}
{Roth} N.,  {Kasen} D.,  {Guillochon} J.,   {Ramirez-Ruiz} E.,  2016, \mn@doi
  [ApJ] {10.3847/0004-637X/827/1/3}, \href
  {https://ui.adsabs.harvard.edu/abs/2016ApJ...827....3R} {827, 3}

\bibitem[\protect\citeauthoryear{{S{\'a}nchez-Janssen}
  et~al.,}{{S{\'a}nchez-Janssen} et~al.}{2019}]{SanchezJanssen_19}
{S{\'a}nchez-Janssen} R.,  et~al., 2019, \mn@doi [ApJ]
  {10.3847/1538-4357/aaf4fd}, \href
  {https://ui.adsabs.harvard.edu/abs/2019ApJ...878...18S} {878, 18}

\bibitem[\protect\citeauthoryear{{Sch{\"o}del}, {Gallego-Cano}, {Dong},
  {Nogueras-Lara}, {Gallego-Calvente}, {Amaro-Seoane}  \&
  {Baumgardt}}{{Sch{\"o}del} et~al.}{2018}]{Schodel_18}
{Sch{\"o}del} R.,  {Gallego-Cano} E.,  {Dong} H.,  {Nogueras-Lara} F.,
  {Gallego-Calvente} A.~T.,  {Amaro-Seoane} P.,   {Baumgardt} H.,  2018,
  \mn@doi [AAP] {10.1051/0004-6361/201730452}, \href
  {https://ui.adsabs.harvard.edu/abs/2018A&A...609A..27S} {609, A27}

\bibitem[\protect\citeauthoryear{{Sersic}}{{Sersic}}{1968}]{Sersic_68}
{Sersic} J.~L.,  1968, {Atlas de Galaxias Australes}

\bibitem[\protect\citeauthoryear{{Stone} \& {Metzger}}{{Stone} \&
  {Metzger}}{2016}]{Stone_16a}
{Stone} N.~C.,  {Metzger} B.~D.,  2016, \mn@doi [MNRAS]
  {10.1093/mnras/stv2281}, \href
  {http://adsabs.harvard.edu/abs/2016MNRAS.455..859S} {455, 859}

\bibitem[\protect\citeauthoryear{{Stone} \& {van Velzen}}{{Stone} \& {van
  Velzen}}{2016}]{Stone_16b}
{Stone} N.~C.,  {van Velzen} S.,  2016, \mn@doi [ApJ]
  {10.3847/2041-8205/825/1/L14}, \href
  {http://adsabs.harvard.edu/abs/2016ApJ...825L..14S} {825, L14}

\bibitem[\protect\citeauthoryear{{Stone}, {Vasiliev}, {Kesden}, {Rossi},
  {Perets}  \& {Amaro-Seoane}}{{Stone} et~al.}{2020}]{Stone2020}
{Stone} N.~C.,  {Vasiliev} E.,  {Kesden} M.,  {Rossi} E.~M.,  {Perets} H.~B.,
  {Amaro-Seoane} P.,  2020, \mn@doi [\ssr] {10.1007/s11214-020-00651-4}, \href
  {https://ui.adsabs.harvard.edu/abs/2020SSRv..216...35S} {216, 35}

\bibitem[\protect\citeauthoryear{{Syer} \& {Ulmer}}{{Syer} \&
  {Ulmer}}{1999}]{Syer_99}
{Syer} D.,  {Ulmer} A.,  1999, \mn@doi [MNRAS]
  {10.1046/j.1365-8711.1999.02445.x}, \href
  {https://ui.adsabs.harvard.edu/#abs/1999MNRAS.306...35S} {306, 35}

\bibitem[\protect\citeauthoryear{{Tadhunter}, {Spence}, {Rose}, {Mullaney}  \&
  {Crowther}}{{Tadhunter} et~al.}{2017}]{Tadhunter_17}
{Tadhunter} C.,  {Spence} R.,  {Rose} M.,  {Mullaney} J.,   {Crowther} P.,
  2017, \mn@doi [Nature Astronomy] {10.1038/s41550-017-0061}, \href
  {http://adsabs.harvard.edu/abs/2017NatAs...1E..61T} {1, 0061}

\bibitem[\protect\citeauthoryear{{Trebitsch}, {Volonteri}, {Dubois}  \&
  {Madau}}{{Trebitsch} et~al.}{2018}]{Trebitsch_18}
{Trebitsch} M.,  {Volonteri} M.,  {Dubois} Y.,   {Madau} P.,  2018, \mn@doi
  [MNRAS] {10.1093/mnras/sty1406}, \href
  {https://ui.adsabs.harvard.edu/abs/2018MNRAS.478.5607T} {478, 5607}

\bibitem[\protect\citeauthoryear{{Vasiliev}}{{Vasiliev}}{2017}]{Vasiliev_17}
{Vasiliev} E.,  2017, \mn@doi [ApJ] {10.3847/1538-4357/aa8cc8}, \href
  {https://ui.adsabs.harvard.edu/#abs/2017ApJ...848...10V} {848, 10}

\bibitem[\protect\citeauthoryear{{Vasiliev}}{{Vasiliev}}{2019}]{Vasiliev_18}
{Vasiliev} E.,  2019, \mn@doi [MNRAS] {10.1093/mnras/sty2672}, \href
  {http://adsabs.harvard.edu/abs/2019MNRAS.482.1525V} {482, 1525}

\bibitem[\protect\citeauthoryear{{Volonteri}, {Lodato}  \&
  {Natarajan}}{{Volonteri} et~al.}{2008}]{Volonteri_08}
{Volonteri} M.,  {Lodato} G.,   {Natarajan} P.,  2008, \mn@doi [MNRAS]
  {10.1111/j.1365-2966.2007.12589.x}, \href
  {https://ui.adsabs.harvard.edu/abs/2008MNRAS.383.1079V} {383, 1079}

\bibitem[\protect\citeauthoryear{{Wang} \& {Merritt}}{{Wang} \&
  {Merritt}}{2004}]{Wang_04}
{Wang} J.,  {Merritt} D.,  2004, \mn@doi [ApJ] {10.1086/379767}, \href
  {http://adsabs.harvard.edu/abs/2004ApJ...600..149W} {600, 149}

\bibitem[\protect\citeauthoryear{{Wevers} et~al.,}{{Wevers}
  et~al.}{2019}]{Wevers_19}
{Wevers} T.,  et~al., 2019, \mn@doi [MNRAS] {10.1093/mnras/stz1602}, \href
  {https://ui.adsabs.harvard.edu/abs/2019MNRAS.487.4136W} {487, 4136}

\bibitem[\protect\citeauthoryear{{van Velzen}}{{van
  Velzen}}{2018}]{VanVelzen_18}
{van Velzen} S.,  2018, \mn@doi [ApJ] {10.3847/1538-4357/aa998e}, \href
  {https://ui.adsabs.harvard.edu/abs/2018ApJ...852...72V} {852, 72}

\bibitem[\protect\citeauthoryear{{van Velzen} \& {Farrar}}{{van Velzen} \&
  {Farrar}}{2014}]{VanVelzen_14}
{van Velzen} S.,  {Farrar} G.~R.,  2014, \mn@doi [ApJ]
  {10.1088/0004-637X/792/1/53}, \href
  {https://ui.adsabs.harvard.edu/abs/2014ApJ...792...53V} {792, 53}

\bibitem[\protect\citeauthoryear{{van Velzen} et~al.,}{{van Velzen}
  et~al.}{2011}]{VanVelzen_11}
{van Velzen} S.,  et~al., 2011, \mn@doi [ApJ] {10.1088/0004-637X/741/2/73},
  \href {http://adsabs.harvard.edu/abs/2011ApJ...741...73V} {741, 73}

\bibitem[\protect\citeauthoryear{{van Velzen} et~al.,}{{van Velzen}
  et~al.}{2020}]{VanVelzen_20}
{van Velzen} S.,  et~al., 2020, arXiv e-prints, \href
  {https://ui.adsabs.harvard.edu/abs/2020arXiv200101409V} {p. arXiv:2001.01409}

\makeatother
\end{thebibliography}
\end{document}